\documentclass[twocolumn,showpacs,aps,prd,amsmath,amssymb,nobibnotes,nofootinbib,floatfix]{revtex4-1}
\newcommand{\beq}{\begin{equation}}
\newcommand{\eeq}{\end{equation}}
\newcommand{\beqn}{\begin{eqnarray}}
\newcommand{\eeqn}{\end{eqnarray}}

\newcommand{\apjl}{Astrophys.\ J.\ Lett.}

\newcommand{\mnras}{Mon. Not. Roy. Astron. Soc.}

\newcommand{\llabel}[1]{\label{#1}}              
\newcommand{\labeq}[2]{ \begin{equation} \llabel{#1}{#2}
\end{NSequation}}

\newcommand\affspc{\vspace{4pt}}
\usepackage{graphicx}
\usepackage{epsf}

\usepackage{graphics,epsfig,placeins,amssymb,pifont}

\usepackage{psfrag}
\usepackage[usenames]{color}
\usepackage{xcolor}
\usepackage{rotating}
\usepackage{hyperref}
\usepackage{ragged2e}

\begin{document}
\title{Binary neutron star mergers: Effects of spin and post-merger
  dynamics}

\author{William E. East${}^1$, Vasileios Paschalidis${}^2$, Frans Pretorius${}^{3,4}$, and Antonios Tsokaros${}^5$}

\affiliation{${}^1$Perimeter Institute for Theoretical Physics, Waterloo, Ontario N2L 2Y5, Canada \affspc}
\affiliation{${}^2$Departments of Astronomy and Physics, University of Arizona, Tucson, Arizona 85721, USA \affspc}
\affiliation{${}^3$Department of Physics, Princeton University, Princeton, New Jersey 08544, USA \affspc} 
\affiliation{${}^4$CIFAR, Cosmology \& Gravity Program, Toronto, Ontario M5G 1Z8, Canada \affspc}
\affiliation{${}^5$Department of Physics, University of Illinois at Urbana-Champaign, Urbana, Illinois 61801, USA \affspc}

\begin{abstract}
Spin can have significant effects on the electromagnetic transients
accompanying binary neutron star mergers. The measurement of spin can
provide important information about binary formation channels. In the
absence of a strong neutron star spin prior, the degeneracy of spin
with other parameters leads to significant uncertainties in their
estimation, in particular limiting the power of gravitational waves to
place tight constraints on the nuclear equation of state.  Thus
detailed studies of highly spinning neutron star mergers are essential
to understand all aspects of multimessenger observation of such
events. We perform a systematic investigation of the impact of neutron
star spin---considering dimensionless spin values up to $a_{\rm
  NS}=0.33$---on the merger of equal mass, quasicircular binary
neutron stars using fully general-relativistic simulations. We find
that the peak frequency of the post-merger gravitational wave signal
is only weakly influenced by the neutron star spin, with cases where
the spin is aligned (antialigned) with the orbital angular momentum
giving slightly lower (higher) values compared to the irrotational
case. We find that the one-arm instability arises in a number of
cases, with some dependence on spin.  Spin has a pronounced impact on
the mass, velocity, and angular distribution of the dynamical ejecta,
and the mass of the disk that remains outside the merger remnant. We
discuss the implications of these findings on anticipated
electromagnetic signals, and on constraints that have been placed on
the equation of state based on multimessenger observations of
GW170817.
\end{abstract}

\maketitle

\section{Introduction}
During the first two observing runs (O1 and O2), the LIGO and Virgo
Scientific Collaborations detected a number of gravitational wave (GW)
signals consistent with the inspiral and merger of binary black
holes~\cite{LIGO_first_direct_GW,Abbott:2016nmj,Abbott:2017vtc,Abbott:2017oio,Abbott:2017gyy,LIGOScientific:2018jsj},
and one GW signal consistent with the inspiral of a binary neutron
star (NS)~\cite{TheLIGOScientific:2017qsa} (GW170817).  GW170817 was
accompanied by a number of observed electromagnetic 
counterparts~\cite{GBM:2017lvd}, including a short gamma-ray
burst~\cite{Monitor:2017mdv}, and an ultraviolet/optical/near-infrared
transient consistent with the radioactive decay of heavy elements
formed in rapidly expanding neutron-rich matter, i.e. a
kilonova~\cite{GBM:2017lvd}. These breakthroughs not only opened a new
window for observing our Universe, but also hold promise that during
the third observing run we could have hundreds of new
events~\cite{LIGO_first_direct_GW}.  Multimessenger observations are
key to solving some long-standing puzzles in fundamental physics and
astrophysics.  For example, observations from GW170817 have already
been used to place new constraints on the behavior of matter at
supernuclear densities (see
e.g.~\cite{TheLIGOScientific:2017qsa,Margalit2017,Shibata:2017xdx,Paschalidis:2017qmb,Bauswein:2017vtn,Ruiz:2017due,Annala:2017llu,Radice:2017lry,Most:2018hfd,Raithel:2018ncd,Abbott:2018exr,Annala:2019puf,Bozzola:2019tit},
and also~\cite{Gandolfi:2019zpj,Raithel:2019uzi} for reviews), and the
sites where heavy elements in the Universe form (see,
e.g.,~\cite{Pian:2017gtc} and references therein). GW170817 has also
been used to independently measure the Hubble
constant~\cite{Abbott:2017xzu}, and to constrain the nature of
gravity~\cite{TheLIGOScientific:2017qsa}. GW170817 ruled out a large
class of modified gravity
theories~\cite{Lombriser:2015sxa,Lombriser:2016yzn,Baker:2017hug,Heisenberg:2017qka,Creminelli:2017sry,Ezquiaga:2017ekz,Sakstein:2017xjx}
by placing a stringent limit on the difference in propagation speed of
GW and electromagnetic waves~\cite{Baker:2017hug}. The wealth of physics extracted
from these first observations has been unprecedented. The fact that
the total observation time of O1 and O2 was only about one year (and
not all detectors were online at all times) and the fact that the
third observing run (O3) will have increased sensitivity, promise that
during O3 there will be many more additional events
and---hopefully---new surprises.

The LIGO/Virgo observation of GW170817 also gave rise to new questions
and puzzles. A particular aspect of the GW170817 observation that is
of interest to this work is that only very weak constraints were
placed on the pre-merger NS spins. Thus, an important question
regarding GW170817 is the following: what was the pre-merger rotation
state of the NSs involved in the event?  This is important for a
number of reasons.  To begin with, uncertainty in the NS spins prior
to merger leads to large uncertainties in other inferred binary
parameters, such as the binary NS total mass and mass ratio, and the
tidal deformability, because of degeneracies in how theses parameters
affect the GW signal~\cite{TheLIGOScientific:2017qsa}. Most studies
placing constraints on the NS equation of state (EOS) assumed the
results from the low-spin prior LIGO analysis of GW170817. While this
assumption is motivated by the fact that the stars in Galactic double
NSs are observed to spin slowly, it nevertheless leads to inferences
about other aspects of the system (including constraints on the
nuclear EOS) that may be biased by our prior observations.  On the
other hand, the observation of non-negligible spin in a merging binary
NS would automatically give revealing information about the formation
channel that gave rise to the binary, perhaps pointing to a
subpopulation of dynamically assembled binaries~\cite{EPP2015}.

Ideally, we want to be able to use these new observations to test our
assumptions and make independent measurements by finding ways orthogonal to the inspiral
GW signal to constrain the NS spins involved in these mergers. For example,
NS spin can potentially affect the lifetime of the merger remnant before it
collapses to a black hole, the amount of disk mass forming outside the remnant,
and/or the mass in dynamical ejecta, all of which affect kilonova signatures
and potential gamma-ray (jet) signatures (see, e.g.,~\cite{Paschalidis:2016agf}
for a recent review). Thus, combining electromagnetic observations and detailed
theoretical modeling of binary NS mergers that accounts for NS spin with GW
observations has the potential for more accurate parameter inference. 

Dynamical spacetime simulations of spinning binary NSs are necessary
to address these issues.
Through these simulations one can
compute the impact of spin on both the GW and electromagnetic
signatures. Simulations of spinning binary NSs are currently
under way by several groups, see,
e.g.,~\cite{Bernuzzi:2013rza,Tacik:2015tja,PhysRevD.95.044045,Dietrich:2015pxa,Dietrich:2017xqb,Ruiz:2019ezy,Most:2019pac,Tsokaros:2019anx}
for quasicircular mergers with constraint-satisfying,
quasiequilibrium initial data, and~\cite{Kastaun2013,Kastaun2015,Kastaun:2016elu}
where the initial data are constraint violating and not in
equilibrium. In~\cite{bauswein2015exploring} simulations were also
presented in the conformal flatness approximation of general
relativity. Finally, we performed studies of eccentric binaries with
spinning NSs, employing constraint-satisfying initial
data in~\cite{EPP2015,PEPS2015,EPPS2016,East:2016zvv}.

Here, we present results from fully relativistic hydrodynamic
simulations of quasiequilibrium binary NSs in
quasicircular orbits with {\it spinning} components. The binary
configurations are of equal mass and equal spin, with the spin
vectors either aligned or antialigned with the orbital angular
momentum. The initial
dimensionless NS spins are in the range~$a_{\rm NS} \in
[-0.13,\,0.33]$, where a positive (negative) sign denotes the corresponding
vectors are aligned (antialigned) with the
orbital angular momentum.  The matter is modeled with different
EOSs which are represented as piecewise polytropes and
cover a range of compactness for a 1.4 $M_\odot$ NS from
$\sim $0.136 to 0.178. We study how spin affects the dynamics of the
merger, the post-merger GW signals, dynamical ejecta, and the merger
remnant disk mass. We also include several nonspinning cases
with different EOSs to illustrate how these effects can be degenerate
with varying EOS, and in order to compare to previous studies.

Our simulations demonstrate that the post-merger peak GW frequency is
only weakly influenced by the NS spin (by about 100 -- 200 Hz). This
is consistent with the results of
Refs.~\cite{Bernuzzi:2013rza,PhysRevD.95.044045} as well as
Ref. \cite{bauswein2015exploring}, the latter using the conformal
flatness approximation to general relativity. We find that aligned
(antialigned) spin cases give slightly lower (higher) values of the
post-merger peak GW frequency when compared to the irrotational
case. In turn, this implies that there is some degree of degeneracy
with the nuclear EOS when inferring the latter from the post-merger
peak GW frequency. We find that the one-arm instability\footnote{We
  note that here and throughout by ``one-arm instability'' we imply
  the existence of a ``one-arm'' ($m=1$) mode that grows out of tiny
  perturbations.} we discovered in eccentric NS mergers (including
those with spin)~\cite{PEPS2015,EPPS2016,East:2016zvv} and studied in
select nonspinning quasicircular
mergers~\cite{Lehner:2016wjg,Radice:2016gym}, also operates in
quasicircular mergers with spin, though the correlation between the
strength of the one-arm mode and the pre-merger spin magnitude is not
strong.  In particular, we find that the strongest one-arm mode
develops for an intermediate value of NS spin that we consider.  The
GWs from post-merger NS oscillations could potentially be detected by
alternative configurations of current observatories optimized for
kilohertz frequencies, as well as third-generation GW
detectors~\cite{Martynov:2019gvu}.

We demonstrate that spin has a substantial impact on the mass,
velocity and angular distribution of dynamical ejecta, and the
subsequent red kilonova signatures. Our results indicate that spins
antialigned with the orbital angular momentum result in more massive
dynamical ejecta, with a non-negligible amount of matter traveling at
speeds near 0.5c. As a result, our study suggests that the radio
signatures of antialigned binary NS (BNS) mergers are expected to be
significantly brighter.  Moreover, our results suggest that
antialigned spin mergers generate brighter red kilonovae than aligned
spin cases, which have smaller dynamical ejecta masses. However, we
find that as the aligned spin increases past a certain value, the
amount of dynamical ejecta increases again. This implies that the
expected red kilonovae should become brighter for higher spin
values---consistent with the fact that as the spin frequency
increases, the star becomes less bound, and hence becomes easier to
dynamically eject more mass. For larger aligned spins, our results
show that the dynamical ejecta are more concentrated near the orbital
plane. Importantly, for dimensionless spins of order 0.2--0.3, our
calculations show that merger remnants have larger disks than
lower-spin cases. Hence, blue kilonovae from such systems are likely
to be brighter as the spin magnitude increases. The fact that as the
spin magnitude increases we obtain heavier disks, implies that the
recent constraints on the binary tidal deformability discussed
in~\cite{Radice:2017lry} may be even weaker than found
in~\cite{Kiuchi:2019lls}, especially since the GW analysis of GW170718
finds that the 90\% confidence interval for the NS dimensionless spins
in GW170817 extends up to 0.6~\cite{Abbott:2018wiz}. Similar
constraints placed on the EOS may need to be revisited, because
related works do not consider the impact of pre-merger spin.

Finally, we compare two simulations that have the same initial
properties, i.e., same total mass, orbital angular frequency, and
equatorial circulation, but one corresponds to corotation and the
other to the corresponding configuration built with the constant
rotational-velocity
formulation~\cite{Tichy:2011gw,Tichy:2012rp,Tsokaros:2018dqs}.  We
find some differences in the post-merger evolution of the two
configurations, but they are broadly consistent with each other.  In
particular, the similarity of the GW emission implies that the bulk of
the matter profiles are quite similar between the two formulations.
Though we do see rather different (but in both cases relativity small)
ejecta masses following merger; this may imply the outer layers in the
stars have slightly different properties between the two formulations.
Given the definition of circulation, and that the velocity
decomposition used in the constant rotational-velocity formulation is
\emph{ad hoc}, this is expected (see discussion in the Appendix of
Ref. \cite{Tsokaros:2018dqs}).

The remainder of this paper is structured as follows. In
Sec.~\ref{numerical_approach}, we briefly describe the numerical methods and
codes we use, including the initial data and the EOSs we treat in
this work. In Sec.~\ref{results_and_discussion}, we present the results from our
numerical simulations and discuss their astrophysical implications. We conclude
in Sec.~\ref{conclusions} with a summary of our main findings and a discussion
of their implications. In Appendix~\ref{sec:covsspi}, we include details of the
comparison of corotation versus constant rotational-velocity methods of
construction spinning binary NSs, and in Appendix~\ref{app:convergence} we give
some details on numerical convergence.  Unless otherwise stated, we use
geometrized units with $G=c=1$.

\section{Numerical approach}
\label{numerical_approach}

We simulate BNS mergers by evolving the Einstein equations
coupled to hydrodynamics using the code described
in Ref.~\cite{code_paper}.  We discretize the Einstein field equations in
the generalized-harmonic formulation using fourth-order accurate
finite differences and time integration.  We model the NS matter as a
perfect fluid, and evolve the general-relativistic Euler equations in
conservative form using the specific high-resolution shock-capturing techniques
detailed in Ref.~\cite{bhns_astro_paper}.

\subsection{Initial conditions}

Our initial data correspond to unmagnetized, quasiequilibrium BNSs in
a quasicircular orbit. For this study, we restrict to binaries
consisting of two identical NSs, modeled by piecewise polytropic EOSs.
We consider cases where each binary companion has an initial
quasilocal dimensionless spin of $a_{\rm NS}:= J_{\rm ql}/(M/2)^2
=-0.13$, $0.08$, $0.17$, $0.25$, and $0.33$, where $J_{\rm ql}$ is the
quasilocal angular momentum of the NS, and $M$ is the
Arnowitt-Deser-Misner (ADM) mass of the binary. For the definition of
the quasilocal angular momentum and subtleties related with it
see~\cite{Tsokaros:2018dqs}.  More specifically, we do not fix the
dimensionless spin, but the circulation, which here is in the range of
$-0.7\mathcal{C}_{\rm cor}$ to $1.9\mathcal{C}_{\rm cor}$, where
$\mathcal{C}_{\rm cor}$ is the circulation of the corotating binary at
that separation~\cite{Tsokaros:2018dqs}. The spinning configurations
are built with the constant rotational-velocity formulation of
Refs.~\cite{Tichy:2011gw,Tichy:2012rp}. Among the spinning
configurations we include two cases that have the same circulation,
ADM mass, and angular velocity, but one is constructed with the
corotation formalism described in Ref.~\cite{Baumgarte:1997eg}, and
the other with the constant rotational-velocity formalism. The
dimensionless spin in both cases is $a_{\rm NS}=0.17$. Given that the
two $a_{\rm NS}=0.17$ configurations are built with different
formulations, it is not \emph{a priori} clear that they describe the same
physical system. Therefore, such comparisons can serve to elucidate
the physics of the constant rotational-velocity formalism for BNS
initial data in a well-understood regime. The details of the
comparison of the simulations with these two configurations are
discussed in Appendix~\ref{sec:covsspi}. All initial data are computed
using the Compact Object CALculator~({\tt
  COCAL})~\cite{Tsokaros:2015fea,Tsokaros:2018dqs}, and their main
properties are listed in Table~\ref{table:NSNS_ID}. The residual
eccentricity for these initial data is $\sim
0.005$~\cite{Tsokaros:2019anx}. For most cases (including all those
with spinning NSs) we use the piecewise polytropic representation of
the ENG EOS~\cite{Engvik:1995gn} from~\cite{Read:2008iy}.  We also
study several nonspinning (irrotational) configurations using the
ENG, 2H, H, and HB EOSs from~\cite{read}.  These cases are included to
give a comparison as to the degree to which the effects of varying
spin are similar to varying EOS, and in particular to be comparable to
the study of eccentric mergers performed in Ref.~\cite{East:2016zvv}.
The radius of a $1.35\ M_\odot$ nonspinning NS with these equations
of state is in the range $[11.6,15.2]$ km.  For the evolutions, we add
an additional thermal component to the pressure, $P_{\rm th} = 0.5
\rho_0 \epsilon_{\rm hot}$ (motivated by~\cite{Bauswein2010}) where
$\epsilon_{\rm hot}$ is the specific energy in excess of that
prescribed by the cold EOS.

%
\begin{center}
  \begin{table*}
    \caption{Properties of the initial BNS configurations.  Listed are
      the EOS, the binary ADM mass $M$, the
      dimensionless ADM angular momentum $J/M^2$, the NS quasilocal
      dimensionless spin parameter $a_{\rm NS}\equiv J_{\rm
        ql}/(M/2)^2$ (aligned ``+'' or antialigned ``$-$'' with orbital
      angular momentum), the approximate rotational period
      $T$~\cite{Tsokaros:2018dqs}, ratio of the coordinate equatorial
      radius toward companion $R_x$ to the coordinate polar radius
      $R_z$, the orbital separation is $D$, corresponding to an
      initial binary angular velocity of $M\Omega$, and circulation
      $\mathcal{C}$. The last column indicates if the binary is a
      spinning ``SP,'' irrotational ``IR'' or corotating ``CO''
      configuration.
      \label{table:NSNS_ID}}
    \begin{tabular}{lccccccccc}
      \hline\hline
          EOS         & $M\,[\rm M_\odot]$  & $J/M^2$   &$a_{\rm NS}$   & $T\,[\rm ms]$ & $R_z/R_x$    & $D\,[\rm km]$ & $M\Omega$ & $\mathcal{C}\,[\rm M_\odot]$ & Spin State  \\  
          \hline
          ENG    &    2.72    &    0.93    &    -0.13    &    4.12    &    0.96    &    41.85    &    0.026    & -3.00   & SP \\
          ENG    &    2.72    &    0.99    &    0.00    &    N/A     &    0.97     &    41.80    &    0.026    &  0.00  & IR \\
          ENG    &    2.72    &    1.02    &    0.08    &    6.14    &    0.97    &    41.77    &    0.026    &  2.00  & SP \\
          ENG    &    2.72    &    1.07    &    0.17    &    3.01    &    0.95    &    41.80    &    0.026    &  4.15  & SP \\
          ENG    &    2.72    &    1.06    &    0.17    &    3.23    &    0.95    &    41.67    &    0.026    & 4.15   & CO \\
          ENG    &    2.72    &    1.11    &    0.25    &    2.13    &    0.92    &    41.89    &    0.026    & 6.00   & SP \\
          ENG    &    2.72    &    1.15    &    0.33    &    1.65    &    0.89    &    42.04    &    0.026    & 8.00  &  SP \\
          2H    &    2.70    &    1.03    &    0.00    &    N/A    &    0.94    &    46.66    &    0.022    &  0.00  & IR \\
          H    &    2.70    &    0.96    &    0.00    &     N/A    &     0.95    &    38.09    &    0.030    & 0.00  &  IR \\
          HB    &    2.70    &    0.99    &    0.00    &    N/A    &    0.97    &    41.62    &    0.026    &  0.00  & IR \\
          \hline\hline
    \end{tabular}
  \end{table*}
\end{center}

\subsection{Diagnostics}
\label{sec:diag}
In order to analyze the simulations, we use several diagnostic quantities.  We
extract the gravitational radiation by evaluating the Newman-Penrose scalar
$\psi_4$ in the wave zone. We decompose this quantity on spheres at large radii
(typically $r=100\ M$) 
into spin $-2$ weighted spherical harmonics with coefficients $C_{\ell
m}$. We also give a frequency domain representation of the GWs by computing the
characteristic strain $h_c=|\tilde{h}|f$ in terms of the Fourier transform of
the strain $\tilde{h}$ and the frequency $f$. 

To characterize the post-merger ejected matter, we use the integrated
rest-mass density $\rho_0$ residing outside some given radius 
\beq
M_0(>r)=\underset{>r\ \ }{\int} \rho_0 u^t \sqrt{-g}\ d^3x 
\eeq 
where $u^t$ is
the $t$ component of the fluid 4-velocity.  Post-merger, part of
the rest mass will become unbound and escape to infinity.  We use the
criteria that $u_t<-1$ and the radial component of the velocity be
positive in flagging fluid elements as unbound. From the value of
$u_t$, we can also determine the distribution of the rest mass $M_0$
over values of the velocity at infinity $v_\infty$.

For comparison, we give a rough estimate of how these measured
properties of the ejecta might translate into observable astrophysical
transients. We do this by use of calculations of such processes
that suggest a rise time for kilonovae light curves
of~\cite{2013ApJ...775...18B}
\begin{equation}
t_{\rm peak}\approx0.3 \left(\frac{M_{\rm
0,u}}{10^{-2}M_{\odot}}\right)^{1/2}\left(\frac{v}{0.2c}\right)^{-1/2} \ \mbox{ d},
\label{tkilonovae}
\end{equation}
measured from the merger, and peak luminosities of
\begin{equation}L\approx 1.6\times 10^{41}\left(\frac{M_{\rm
0,u}}{10^{-2}M_{\odot}}\right)^{1/2}\left(\frac{v}{0.2c}\right)^{1/2}\mbox{
   erg $s^{-1}$}\label{Lkilonovae}.
\end{equation}
Here, $M_{\rm 0,u}$ and $v$ are the rest mass and characteristic
velocity of the unbound ejecta. Note that Eqs.~\eqref{tkilonovae}
and~\eqref{Lkilonovae} depend on the opacity of the ejecta, which
depends on their composition. Here we scale the equations such that
they are in agreement with the results of~\cite{2013ApJ...775...18B}.
It is possible that ejecta from mergers with spinning neutron stars
and different equations of state have slightly different composition
than nonspinning ones, thus Eqs.~\eqref{tkilonovae}
and~\eqref{Lkilonovae} are used to provide simple estimates and to
understand any trends. Detailed radiative transfer calculations of
ejected matter are necessary for robust kilonova calculations, see
e.g.~\cite{Kawaguchi_2018}.

Typical unbound ejecta masses from BNS mergers are of order
$10^{-3}$--$10^{-2}M_\odot$. However, the observed kilonova
accompanying GW170817 has been explained by invoking ejecta masses of
order
$0.025-0.05M_\odot$~\cite{Coulter:2017wya,Drout:2017ijr,Shappee:2017zly,Kasliwal:2017ngb,
  Tanaka:2017qxj,Arcavi:2017xiz,Pian:2017gtc,Smartt:2017fuw,Soares-Santos:2017lru,
  Nicholl:2017ahq,Cowperthwaite:2017dyu}, and hence significantly
larger than the dynamical ejecta masses. Moreover, the kilonova
associated with GW170817 seems to require at least two components to
explain the observed color evolution: one component accounting for the
red kilonova (explained by high-opacity, lanthanide-rich outflows
associated with low-electron fraction dynamical ejecta), and one
component accounting for the blue kilonova (explained by low-opacity,
lanthanide-poor outflows associated with high-electron fraction disk
wind material). Therefore, it is not likely that dynamical ejecta from
BNSs can, by itself, explain such bright electromagnetic signatures.

Recent
studies~\cite{Fernandez:2013tya,Metzger:2014ila,Perego:2014fma,Just2015MNRAS.448..541J,Wu:2016pnw,Lippuner:2017bfm,Siegel:2017nub}
suggest that a significant fraction of the mass of the disk that forms
around the BNS merger remnant becomes unbound because of viscous,
neutrino, and magnetic field processes. Thus, a disk mass of order
$0.05-0.1M_\odot$ can explain the kilonova that accompanied
GW170817. For these reasons, we also estimate the disk rest mass as
the bound mass outside the remnant black hole. In those cases where a
longer-lived massive NS forms following merger, we define the disk
rest mass as the bound rest mass outside a radius of $\approx30$ km,
which is where, in our simulations, the rest-mass density roughly
drops by 3 orders of magnitude from the maximum rest-mass density
of the massive NS remnant. We point out that the definition of disk
mass is not unambiguous (see e.g.~\cite{PhysRevD.98.043015}). However,
if we define the remnant star surface as the isosurface with rest-mass
density 2 orders of magnitude smaller than the maximum value of the
rest-mass density, then the stellar radii in all of our
simulations are all smaller than 20 km. In this sense, the disk masses we
compute for radii $> 30$ km are a lower limit.

Another electromagnetic transient associated with material ejected in
compact object mergers is radio emission when this material sweeps the
interstellar medium~\citep{2011Natur.478...82N}.  These signals
typically peak on time scales~\citep{2011Natur.478...82N}
\begin{equation}
t_F \approx 6 \left(\frac{E_{\rm kin}}{10^{51}\mbox{\ 
   erg}}\right)^{1/3}\left(\frac{n_0}{0.1\ \mbox{cm}^{-3}}\right)^{-1/3}\left(\frac{v}{0.3c}\right)^{-5/3} \rm yr
\label{EjectaISMtime}
\end{equation}
with brightness
\begin{eqnarray} 
 F(\nu_{\rm obs}) &\approx& 0.6\left(\frac{E_{\rm kin}}{10^{51}\mbox{
     erg}}\right)\left(\frac{n_0}{0.1 \ {\rm
     cm}^{-3}}\right)^{7/8} \label{Fnu} \\ &&
 \left(\frac{v}{0.3c}\right)^{11/4}\left(\frac{\nu_{\rm obs}}{{\ \rm
     GHz}}\right)^{-3/4}\left(\frac{d}{100 \ {\rm Mpc}}\right)^{-2}
 \mbox{ mJy}.\nonumber
\end{eqnarray} 
In the equations above, $E_{\rm kin}$ is the kinetic energy of the
ejecta, $\nu_{\rm obs}$ is the observation frequency, $d$ the distance
to source, and we use a fiducial value of 
$n_0\sim0.1$ cm$^{-3}$
for the interstellar medium density.

In cases where a massive NS remnant
forms post-merger, we also characterize the matter using several quantities.  We
decompose the density distribution into different azimuthal modes \beq
C_m = \int \rho_0u^t\sqrt{-g} e^{im\phi}\ d^3x, \eeq where $\phi$ is
the coordinate azimuthal angle and $m$ is an integer, which 
is useful for monitoring the onset of
shear/nonaxisymmetric instabilities. In particular, for all the configurations
considered here, $C_m$ is initially zero for odd $m$, and becomes nonzero
through such instabilities.

For these massive NS remnants, we also examine the rotational profile
of the post-merger star using an azimuthal average of the angular
velocity $\Omega=u^\phi/u^t$ over fixed values of the cylindrical
coordinate radius $\varpi=\sqrt{x^2+y^2}$, where we compute $u^\phi$,
and $\varpi$ in the initial center-of-mass frame of the binary, i.e.,
the origin of the coordinate system.

\subsection{Resolution}
For all simulations, we utilize six levels of adaptive mesh refinement
each with $2:1$ refinement ratio. For most of the results presented
here, we use resolution with $193^3$ points on the base level, a
resolution of $dx\approx 0.05 M \approx 0.2{\rm \ km}$ on the finest level,
and a resolution of $dx\approx 1.6 M\approx 1/(47\ \rm kHz)$ in the GW extraction zone.  
The grid structure is dynamically adjusted during the evolution based on 
truncation error estimates for the metric functions, while ensuring that the
finest level always covers the star(s). See Ref.~\cite{code_paper} for more details.
To establish
convergence, and estimate truncation errors, we also run two cases at
$4/3\times$ and $2\times$ our canonical resolution. Results of the
convergence study are presented in Appendix~\ref{app:convergence}.

\section{Results and discussion}
\label{results_and_discussion}
With the initial parameters used here, the BNSs undergo
$\sim4$--6 orbits before merging. Of the mergers we study here, we
find that---with the exception of the one case using the HB EOS---all
produce a long lived massive NS remnant that does not
collapse on the timescales of our simulations (up to $\sim30$ ms
post-merger).  In the following, we characterize the gravitational
wave signals, post-merger remnants, and unbound material from these
cases with different EOSs and spins.

\subsection{Gravitational waves\label{sec:gw}} 
\begin{figure}
\begin{center}
	\includegraphics[width = 3.5in]{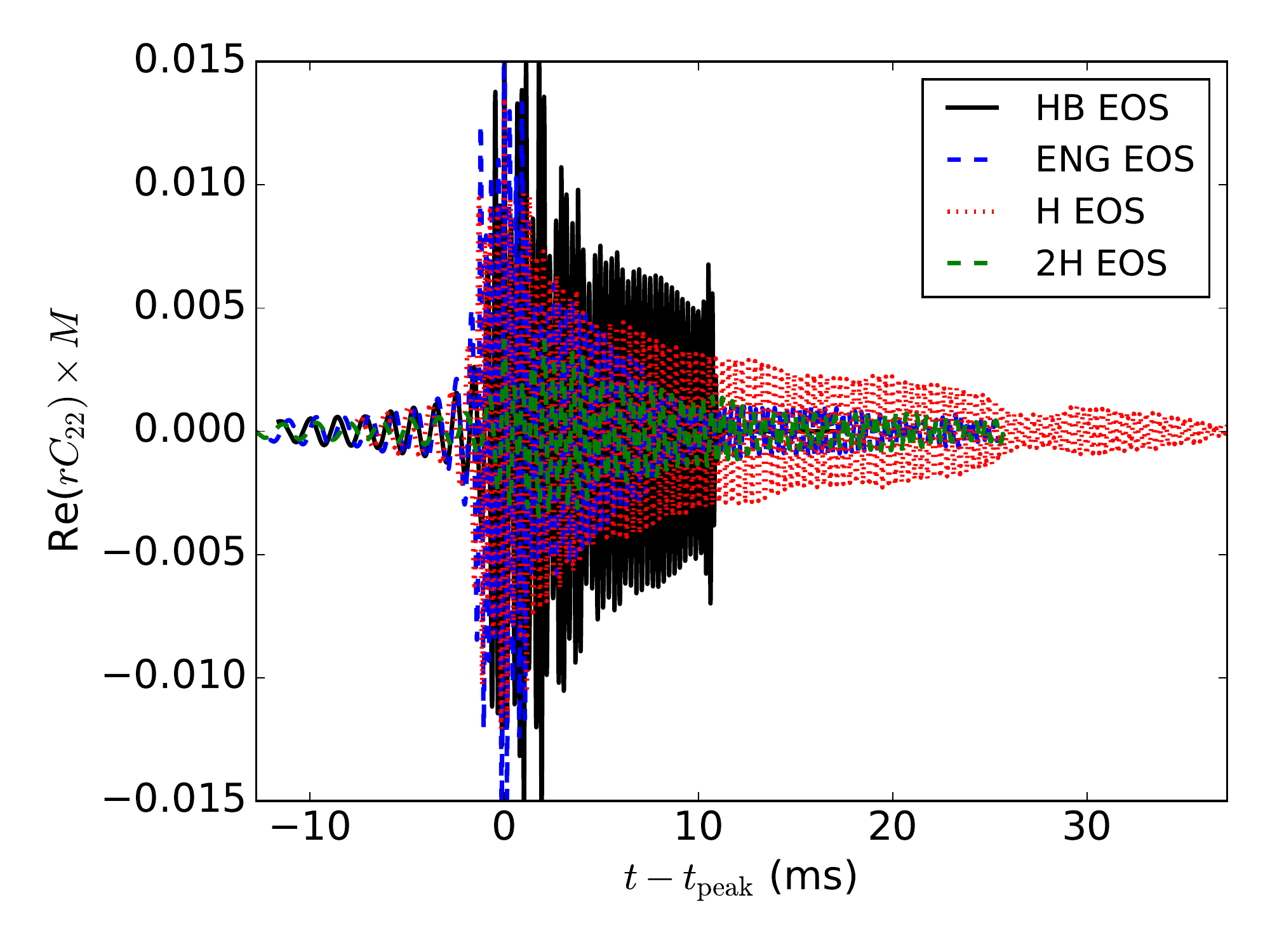}
\caption{
	The GW signal ($\ell=m=2$ component of
	$\psi_4$ multiplied by the extraction radius $r$) from nonspinning mergers with various EOSs. The
	curves have been aligned in time and phase at peak. The HB EOS
        curve ends once a black hole (BH) forms; the other cases did not collapse
        to BHs during the time of their respective simulations.
 } \label{fig:gw_eos}
\end{center}
\end{figure}

For the mergers of nonspinning NSs with different EOSs, the post-merger GW
amplitude is larger for softer EOSs that give rise to more compact NSs. 
This is illustrated in Fig.~\ref{fig:gw_eos} which shows the 
dominant $\ell=m=2$ component of $\psi_4$. For the softest EOS considered,
the HB, the BNS remnant collapses after $\sim10$ ms producing a black hole.

\begin{figure*}
\begin{center}
	\includegraphics[width = 3.5in]{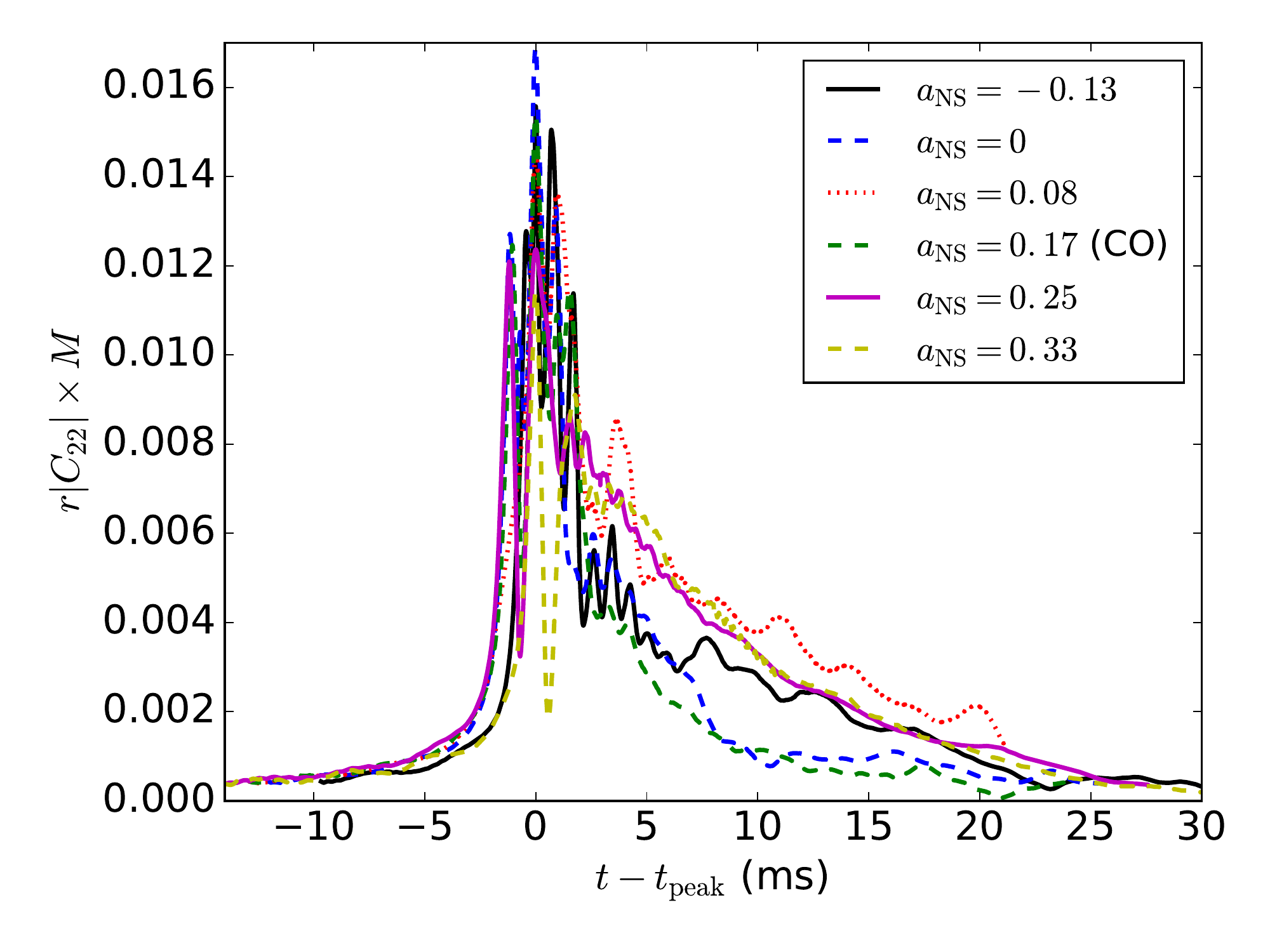}
	\includegraphics[width = 3.5in]{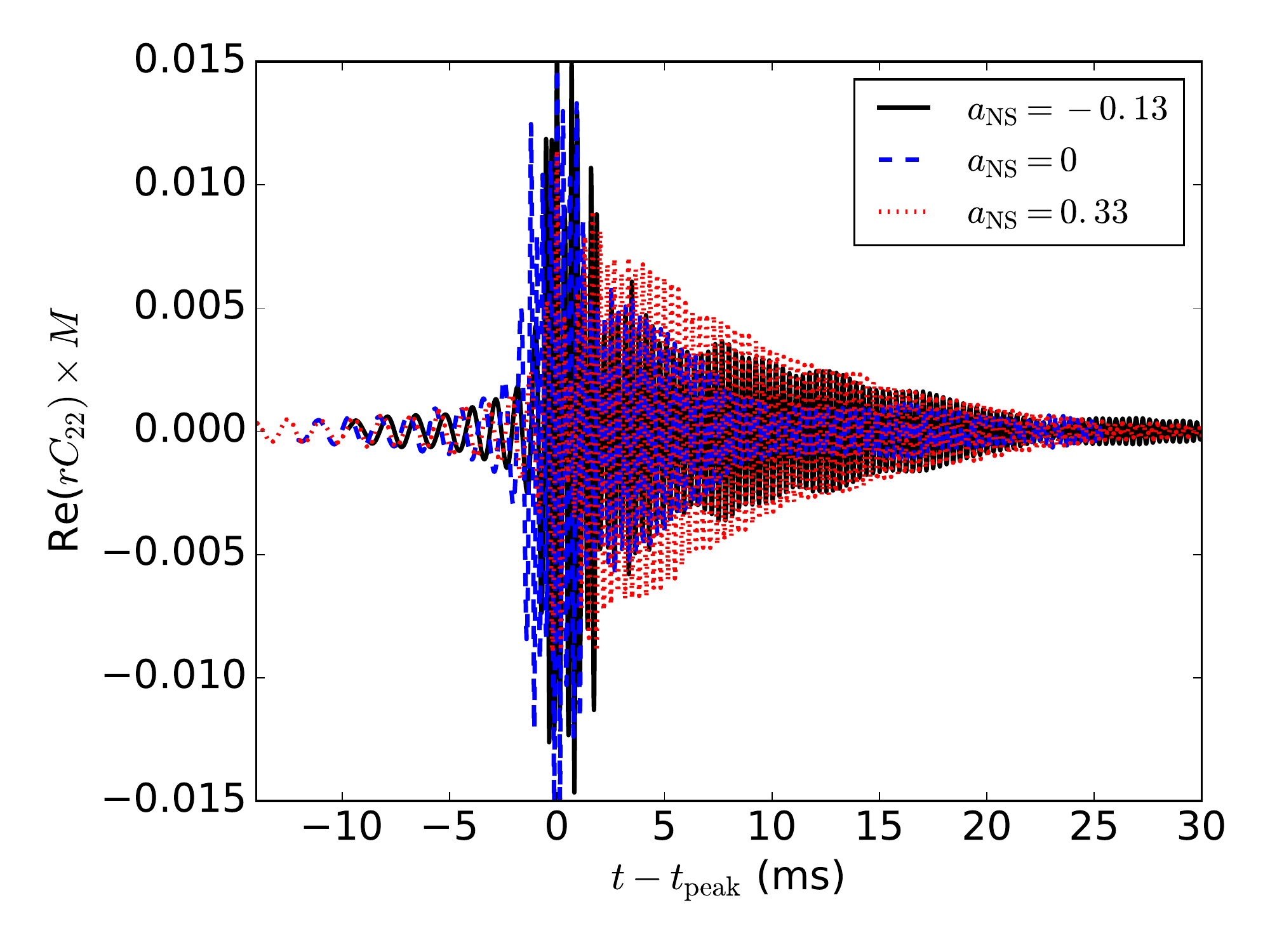}
\caption{
	The GW signal ($\ell=m=2$ component of
	$\psi_4$ multiplied by the extraction radius $r$) from mergers with the ENG EOS and various
	values of NS spin. 
    The left panel shows the magnitude of $\psi_4$, while the right panel shows the real part
    of $\psi_4$ for select cases.
    The curves have been aligned in time and (for the right panel) phase at peak. 
        None of these cases collapsed to BHs during the time of their respective simulations.
 } \label{fig:gw_eng}
\end{center}
\end{figure*}

We also show the GWs for the ENG EOS and various values of NS spin in
Fig.~\ref{fig:gw_eng}. We observe that higher prograde spin ($a_{\rm
  NS} \gtrsim 0.17$) results in smoother post-merger GWs that do not
exhibit ``beat'' oscillations as in the lower-spin cases. Moreover,
the post-merger GW amplitude in nonzero spin cases decays more slowly
in the first 20 ms than in the irrotational case, implying a stronger
post-merger GW signal when spin is considered. This result is further
supported by our resolution study in Appendix~\ref{app:convergence}.

\begin{figure*}
\begin{center}
	\includegraphics[width = 3.5in]{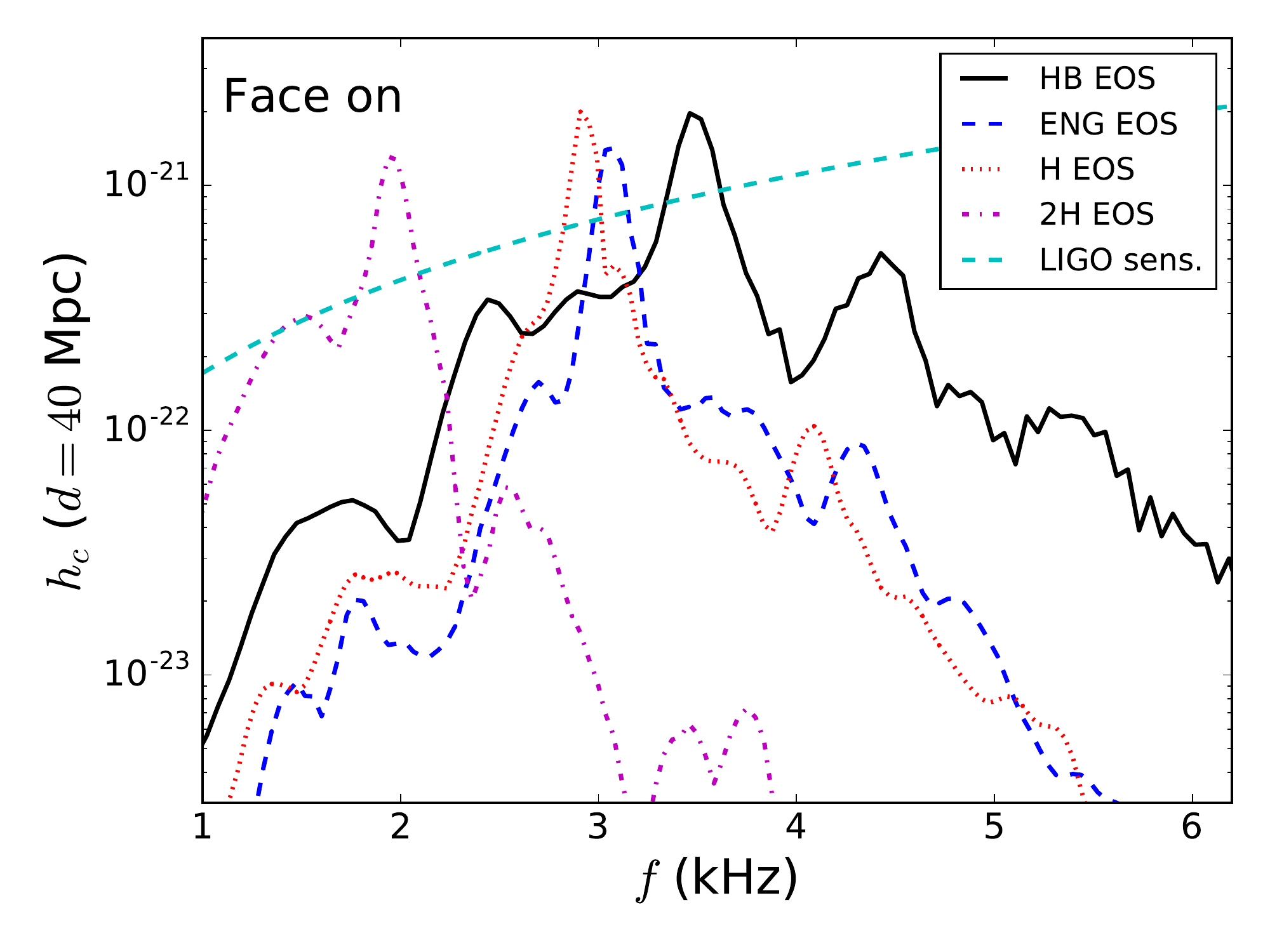}
	\includegraphics[width = 3.5in]{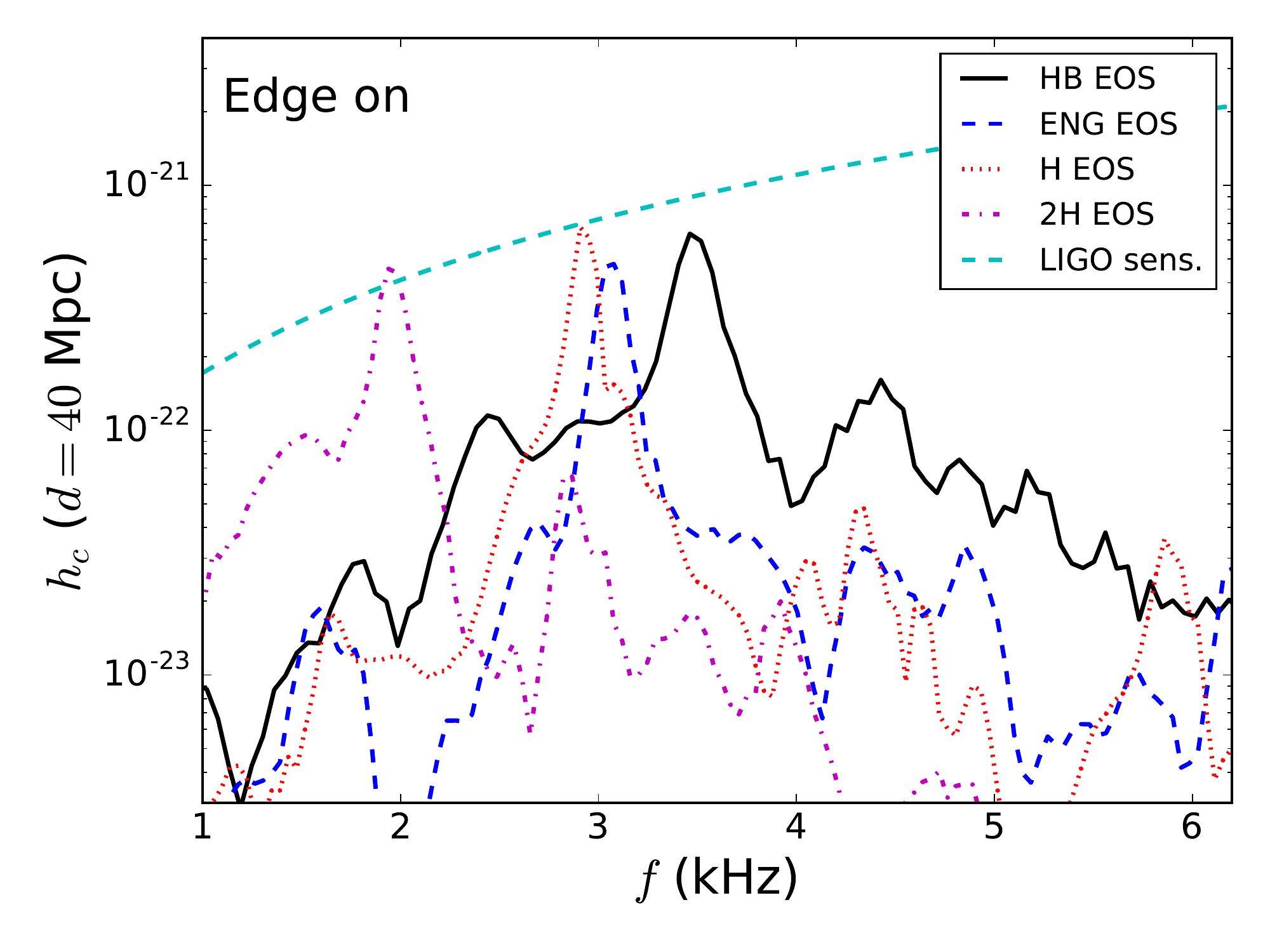}
\caption{ The characteristic strain as a function of frequency for the
  post-merger GW signal for irrotational cases with different EOSs
  (computed in a $\approx 20$ ms window following the peak GW
  luminosity signal) as seen by an observer oriented face-on (left) or
  edge-on (right) with respect to the orbital plane. All $2\leq \ell \leq 6$ modes
  are used in the plot.  The $m=1$, and $m=-1$ modes are those driving
  the peak at $f\sim 1.6\rm kHz$ in the right panel. The case with the
  HB EOS collapses to a BH during this time.  } \label{fig:fgw_eos}
\end{center}
\end{figure*}

\begin{figure*}
\begin{center}
	\includegraphics[width = 3.5in]{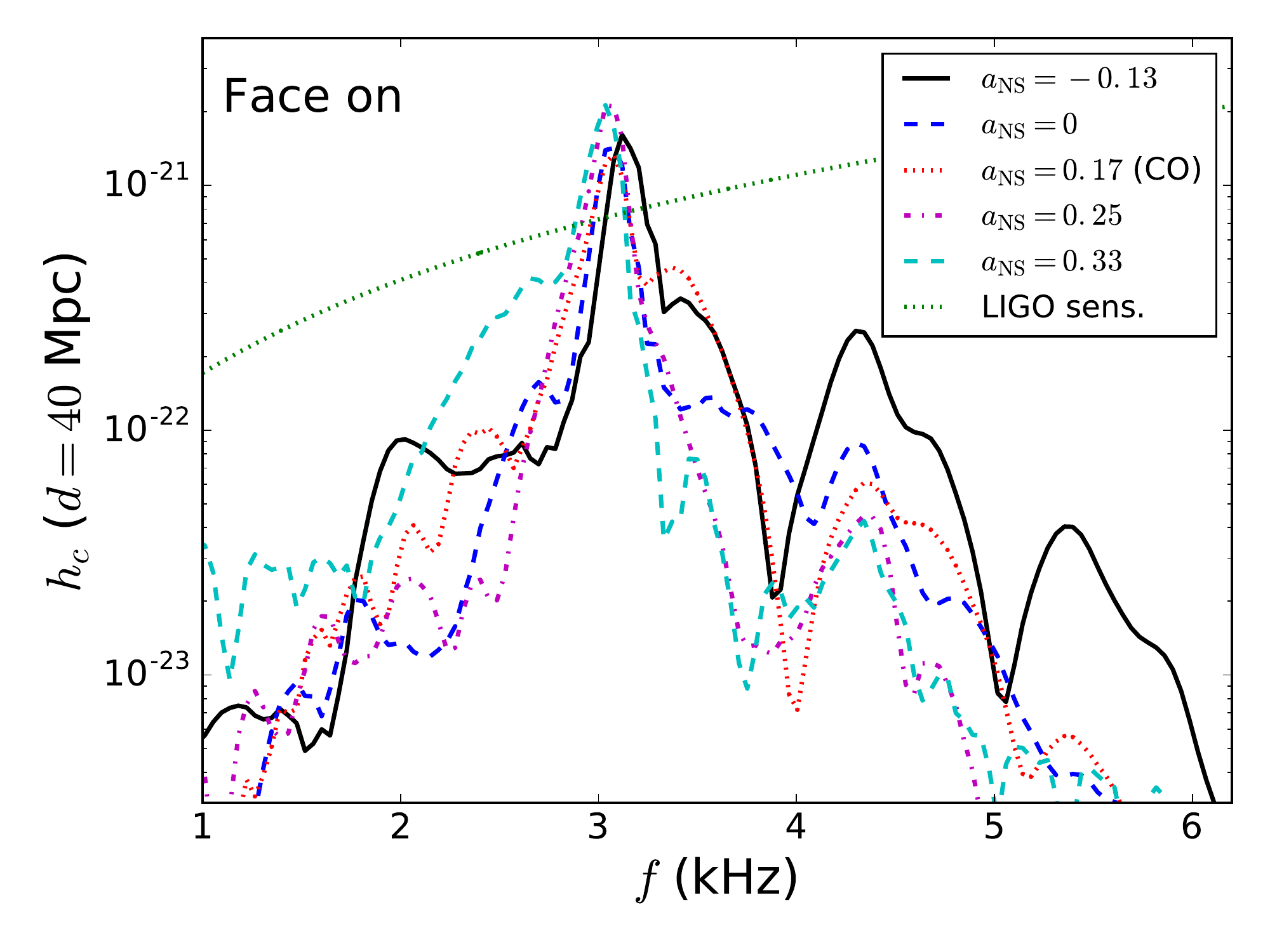}
	\includegraphics[width = 3.5in]{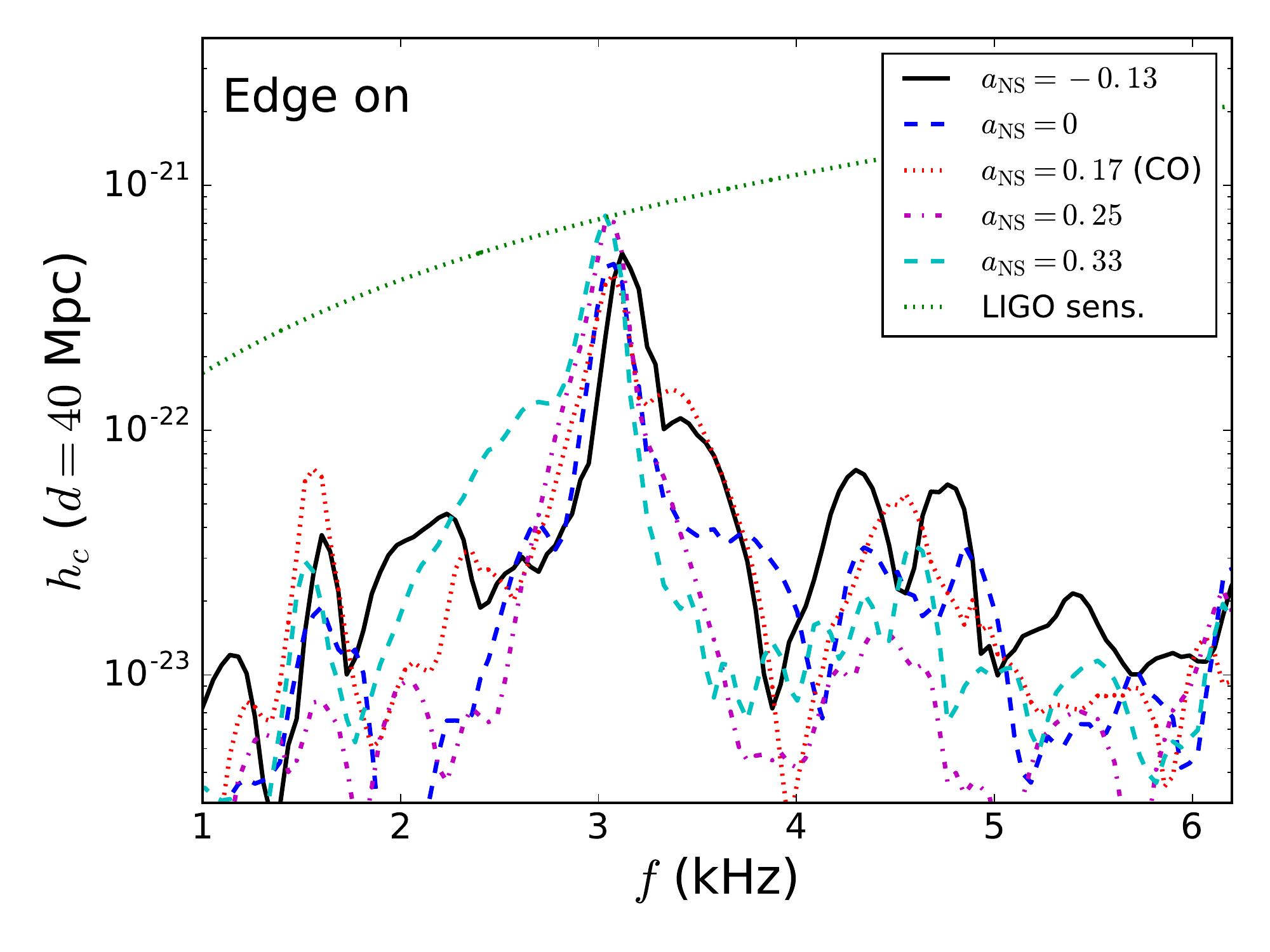}
\caption{ The characteristic strain as a function of frequency for the
  post-merger GW signal for cases with the ENG EOS and various spins
  (computed in a $\approx 20$ ms window following the peak GW
  luminosity signal) as seen by an observer oriented face-on (left) or
  edge-on (right) with respect to the orbital plane. All $2\leq \ell \leq 2$ modes
  are used in the plot.  The $m=1$, and $m=-1$ modes are those driving
  the peak at $f\sim 1.6\rm kHz$ in the right panel
} \label{fig:fgw_eng}
\end{center}
\end{figure*}

The post-merger GW signal is strongly peaked in a narrow frequency
range, as more clearly illustrated in the frequency domain
decompositions shown in Figs.~\ref{fig:fgw_eos} and \ref{fig:fgw_eng}. 
Figure~\ref{fig:fgw_eos} shows how the peak frequency noticeably shifts
to lower frequencies as stiffer EOSs are considered for irrotational binaries.
The peak frequency is
also weakly influenced by the NS spin (by about $100$--$200$ Hz), as evident
in Fig.~\ref{fig:fgw_eng}, with
the aligned (antialigned) spin cases giving slightly lower (higher)
values compared to the nonspinning case. This implies that when
considering the post-merger peak GW frequency as a means for
constraining the nuclear EOS, there will be some
degeneracy between the EOS and the spin.  As noted above, 
the cases with
higher spin magnitude also have somewhat more power at the peak
frequency. There are also smaller
components in the GW signal at other frequencies, as can be seen in
Figs.~\ref{fig:gw_eng}. 

From the left panel of Fig.~\ref{fig:fgw_eng}, it is evident that the
antialigned case has the most GW power in the higher frequency subdominant
peaks, and that this power decreases strongly as the aligned NS spin is
increased, though the frequency at which the peaks occur is relatively
insensitive to the spin. However, such features at very high frequencies
($\gtrsim4$ kHz) will be difficult to observe in the near future.  At lower
frequencies, a prominent feature in the edge-on view (right panel of
Fig.~\ref{fig:fgw_eng}) is the peak close to $1.6$ kHz, which corresponds to
the frequency of the one-arm mode as found in studies of eccentric BNS
mergers~\cite{PEPS2015,EPPS2016,East:2016zvv} and quasicircular
mergers~\cite{Lehner:2016wjg,Radice:2016gym}.  We discuss the post-merger
dynamics of the one-arm mode in more detail in the following section, and here
we focus on its detectability in the GW spectrum.

As pointed out in~\cite{PEPS2015,EPPS2016,East:2016zvv}, the frequency
of the one-arm mode occurs at half the peak frequency, and hence where
LIGO/Virgo are more sensitive.  Moreover, the massive NSs formed after
merger emit almost monochromatic GWs at the one-arm mode frequency. We
conjecture that this monochromatic emission can persist for much
longer than the $20$ ms windows used for generating
Fig.~\ref{fig:fgw_eng}, building more power. The largest amplitude
among the different cases occurs for spinning cases, and in particular
for the corotating configuration with the ENG EOS. We can estimate the
approximate strength of the long-lived gravitational signal in this
case as follows.  If we assume that the source is observed on edge and
that the $m=1$ mode has constant frequency and amplitude, the
signal-to-noise ratio (SNR) for the $m=1$ mode can be estimated via
Eq. (81) of~\cite{Jaranowski1998} (see also~\cite{Lehner:2016wjg}) and
approximating the $\ell=2$, $m=1$ mode GW strain as $h_{21}\sim
C_{21}/(2\pi f_{m=1})^2$
\begin{widetext}
\begin{equation}\label{ligo_snr}
  {\rm SNR}_{\rm LIGO} \approx 3 \left(\frac{7\times 10^{-24}
    {\rm
    \  Hz}^{-1/2}}{\sqrt{S_n(f_{m=1})}}\right)\left(\frac{C_{21}r M}{10^{-4}}\right)\left(\frac{1.6
    {\rm \ kHz}}{f_{m=1}}\right)^{2}\left(\frac{T_{m=1}}{100 {\rm
     \ ms}}\right)^{1/2}\left(\frac{10{\rm \ Mpc}}{r}\right)
\end{equation}
\end{widetext}
for the LIGO zero-detuned high power configuration at design sensitivity.
Here, $S_n(f_{m=1})$ is the detector noise spectral density at the frequency of
the one-arm mode, and we adopt a mode lifetime of $T_{m=1}=100$ ms (order of
magnitude consistent with some of our simulations) and distance to the source
$r=10$ Mpc.  Such high frequency GW signals will only be seen by LIGO for very
close events.  However, the prospects for third-generation detectors are
better: e.g.  the Einstein Telescope ET-D configuration~\cite{ET} would have
$10\times$ the SNR, and there are proposals for obtaining comparable
sensitivities in the kilohertz regime by modifying existing ground-based
detectors~\cite{Martynov:2019gvu}.  The lifetime of the one-arm mode may be
considerably extended in the cases of initial prograde NS spin due to the
additional centrifugal support by the increased total angular momentum which
can extend the lifetime of the remnant.  Since the one-arm mode decays very
slowly, if the remnant survives collapse to a BH for longer times, the mode may
survive for longer times. Note that, as pointed out in~\cite{Lehner:2016wjg},
detection of the inspiral GW from a BNS will substantially lower the SNR
requirements to claim a detection of a post-merger $l=2$, $m=1$ GW mode. In
addition, coherent stacking data analysis methods can be adopted to boost the
signal from the detection of a number of sources~\cite{Yang:2017xlf} (see also
~\cite{Bose:2017jvk}).

\subsection{Properties of the merger remnants}
For the NS merger with the HB EOS, the post-merger remnant collapses
to a BH with $M_{\rm BH}\approx 0.94\ M$ and a dimensionless spin of
$a_{\rm BH}\approx 0.6$ during the simulation. For all other cases,
the remnant star lasts the entire extent of our simulations. In the
following we discuss the angular velocity profiles and
nonaxisymmetric density modes that are excited in these remnants.

\subsubsection{Angular velocity}
The remnant stars in our simulations settle into a slowly varying
pattern of differential rotation. In recent studies, it has been found
that the angular velocity profile ($\Omega$ versus the cylindrical
radius $\varpi$) in massive BNS merger remnants formed from initially
irrotational configurations appears to have an approximately universal
shape independent of initial data, gauge conditions, equations of
state, and initial spin (when starting with constraint-violating and
nonequilibrium initial data). This angular velocity profile is
constant near the center of the star, then as $\varpi$ increases,
$\Omega$ increases until it reaches a maximum value, after which
$\Omega$ falls off with
$\varpi$~\cite{Kastaun2015,Kastaun:2016yaf,Kastaun:2016yaf,Kastaun:2016elu,2016arXiv160403445E,Hanauske:2016gia,PhysRevD.96.043004,Ciolfi:2017uak,PhysRevD.98.043015,Ciolfi:2019fie}. Here
we test whether this profile depends on the initial NS spin, when
starting with constraint-satisfying and equilibrium initial data.

\begin{figure}
\begin{center}
  \includegraphics[width = 0.49\textwidth]{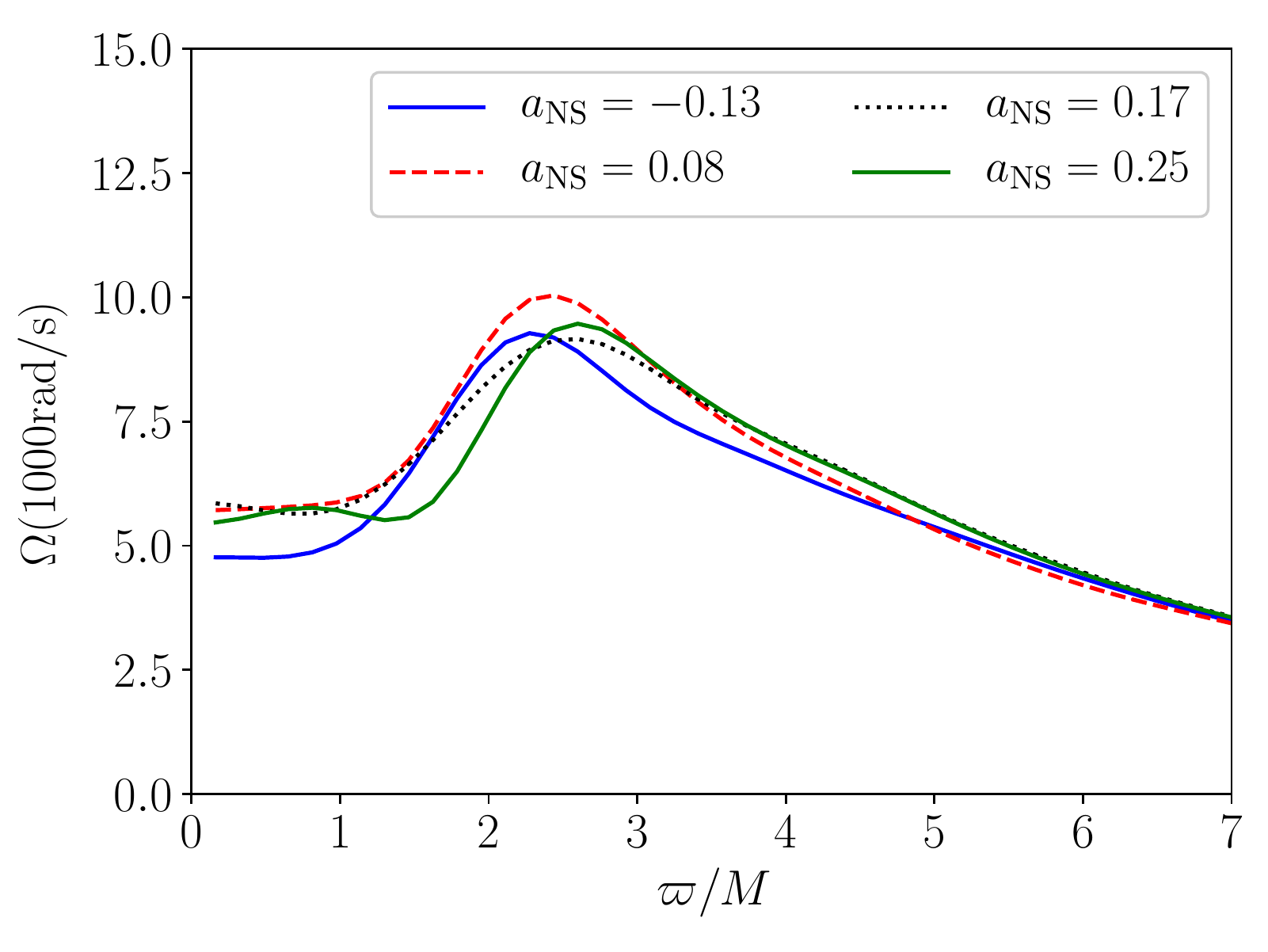}
        \caption{ The angular velocity versus cylindrical coordinate
          radius on the equator of the massive NS remnant for select
          cases with the ENG EOS. The $a_{\rm NS}=0.17$ case shown
          corresponds to the corotating initial data.
        } \label{fig:hmns_rot}
\end{center}
\end{figure}

In Fig.~\ref{fig:hmns_rot}, we show the late-time azimuthally averaged
angular velocity profiles of cases with different initial spins. The
late-time rotational profiles appear to be relatively insensitive to
the initial spin. However, we point out that these calculations are
not gauge invariant, nor is the initial center of mass the true center
of rotation of these configurations. Moreover, within several tens of
ms from merger, magnetic fields can brake the differential rotation,
bringing the core to a near uniformly rotating
state~\cite{Ruiz:2019ezy}. Thus, these $\Omega$ profiles are not
expected to be long-lived, even if the remnant survives for long
times.

\subsubsection{Nonaxisymmetric instabilities}
\begin{figure*}
\begin{center}
	\includegraphics[width = 3.5in]{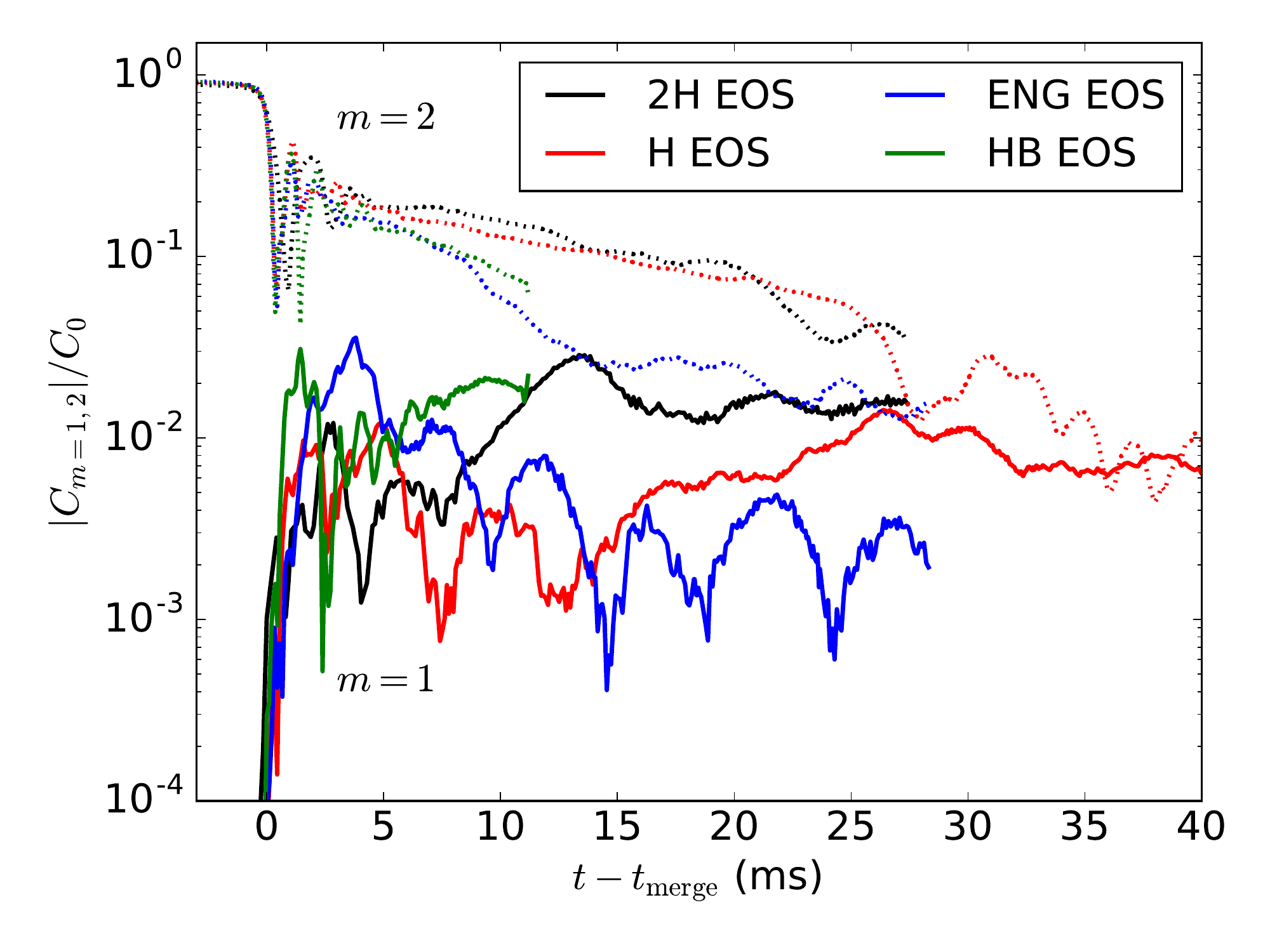}
	\includegraphics[width = 3.5in]{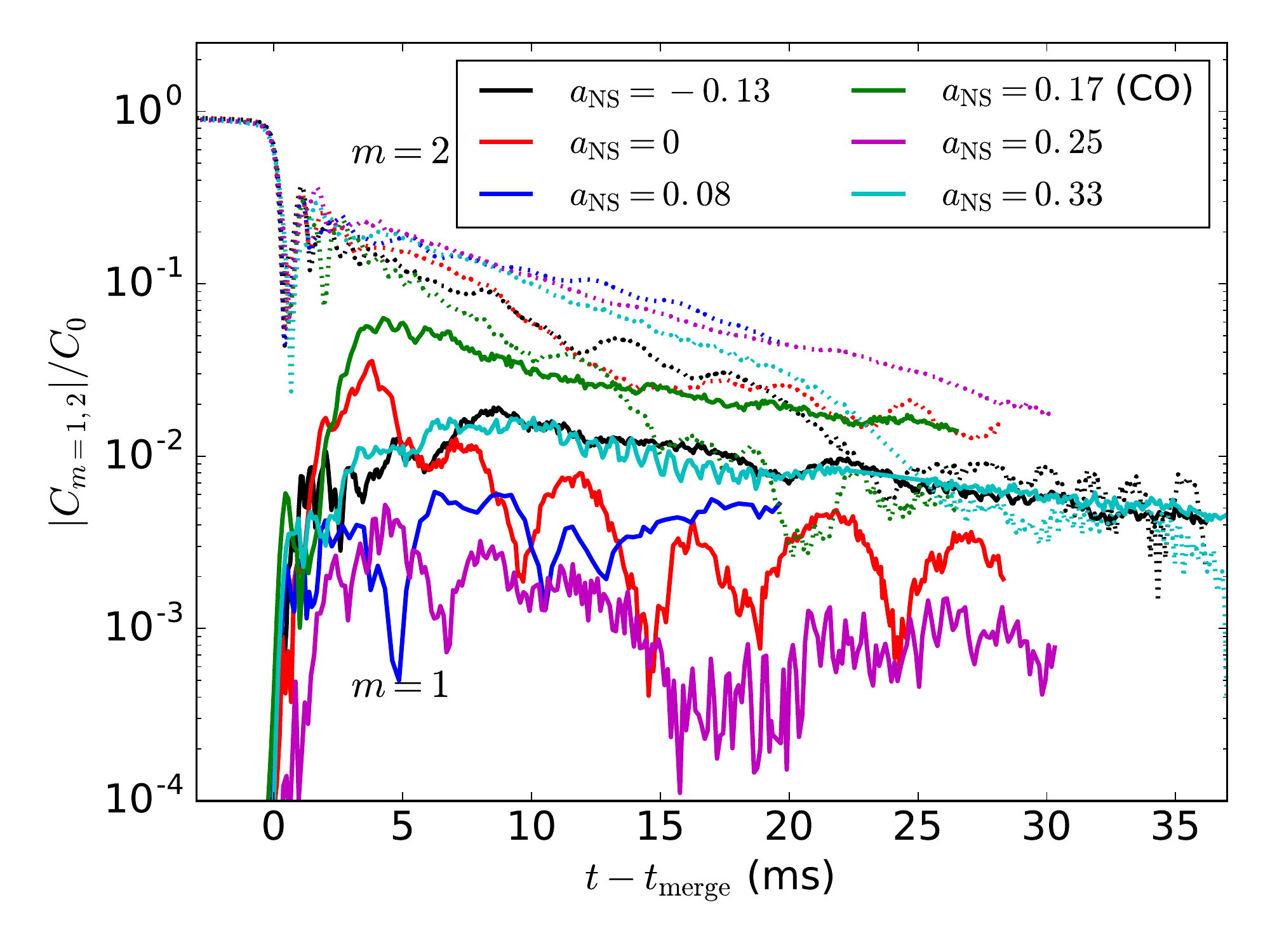}
\caption{
    The magnitude of the $m=1$ and $m=2$ azimuthal density modes as a function
    of time for nonspinning cases with various EOSs (left) and for the ENG EOS
    with various values of NS spin (right). The curves for the HB EOS end at
    the time of BH formation; the curves for the other cases end when their
    respective simulations were stopped.
 } \label{fig:azi_dens}
\end{center}
\end{figure*}

We also quantify how the azimuthal matter distribution of the
post-merger remnant evolves with time.  In Fig.~\ref{fig:azi_dens}, we
show the magnitude of the $m=1$ and 2 azimuthal density modes. Since
all cases considered here initially have identical binary
constituents, the $m=1$ component is essentially zero pre-merger, but
is generated by shear instabilities during the merger.  All cases
excite the $m=1$ mode, but for several cases---namely the irrotational
2H and H EOSs, as well as the ENG $a_{\rm NS}=-0.13$, $a_{\rm
  NS}=0.17$ CO (and SP to a slightly lesser extent; see
Appendix~\ref{sec:covsspi}), and $a_{\rm NS}=0.33$---the combination
of the persistence of the merger-generated $m=1$ mode combined with
the more rapid decay of the $m=2$ density perturbation means that the
former is comparable or greater in magnitude to the latter at late
times. For the ENG $a_{\rm NS}=-0.13$, IR ($a_{\rm NS}= 0$), CO
$a_{\rm NS}=0.17$, and $a_{\rm NS}=0.33$ cases, we have also Fourier
transformed the $m=1$ density modes shown in Fig.~\ref{fig:azi_dens},
and found that the peak $m=1$ frequencies occur at $1656$ Hz, $1655$ Hz,
$1593$ Hz, and $1584$ Hz, respectively. These frequencies agree well
with the frequencies of the corresponding peaks at $\sim 1600$ Hz in
the GW spectrum shown on the right panel in Fig.~\ref{fig:fgw_eng},
indicating that it is the $m=1$ mode that is driving that peak.  In
addition, these frequencies demonstrate that as the pre-merger NS spin
increases the additional angular momentum shifts the one-arm mode
frequency to smaller values.

Therefore, our study suggests that the $m=1$ instability, as found in
eccentric mergers~\cite{PEPS2015,EPPS2016,East:2016zvv} and certain
irrotational quasicircular binary mergers including unequal mass
cases~\cite{Radice:2016gym,Lehner:2016wjg}, can also arise in equal
mass quasicircular BNS mergers including spin, and leaves a similar
imprint on the GW spectrum. However, we do not find a particular trend
of the mode amplitude with initial spin.

We note that the growth of the $m=1$ instability can be tracked in a
gauge invariant way through the mode decomposition of the
GWs. In cases we study here, the amplitude of $m=1$
GW modes grows rapidly during merger and saturates
post-merger, as we have found in previous studies (see Fig.~13
in Ref.~\cite{EPPS2016} and Fig.~2 in Ref.~\cite{East:2016zvv}).

\subsection{Post-merger matter distribution and electromagnetic counterparts}
\begin{figure*}
\begin{center}
	\includegraphics[width = 3.5in]{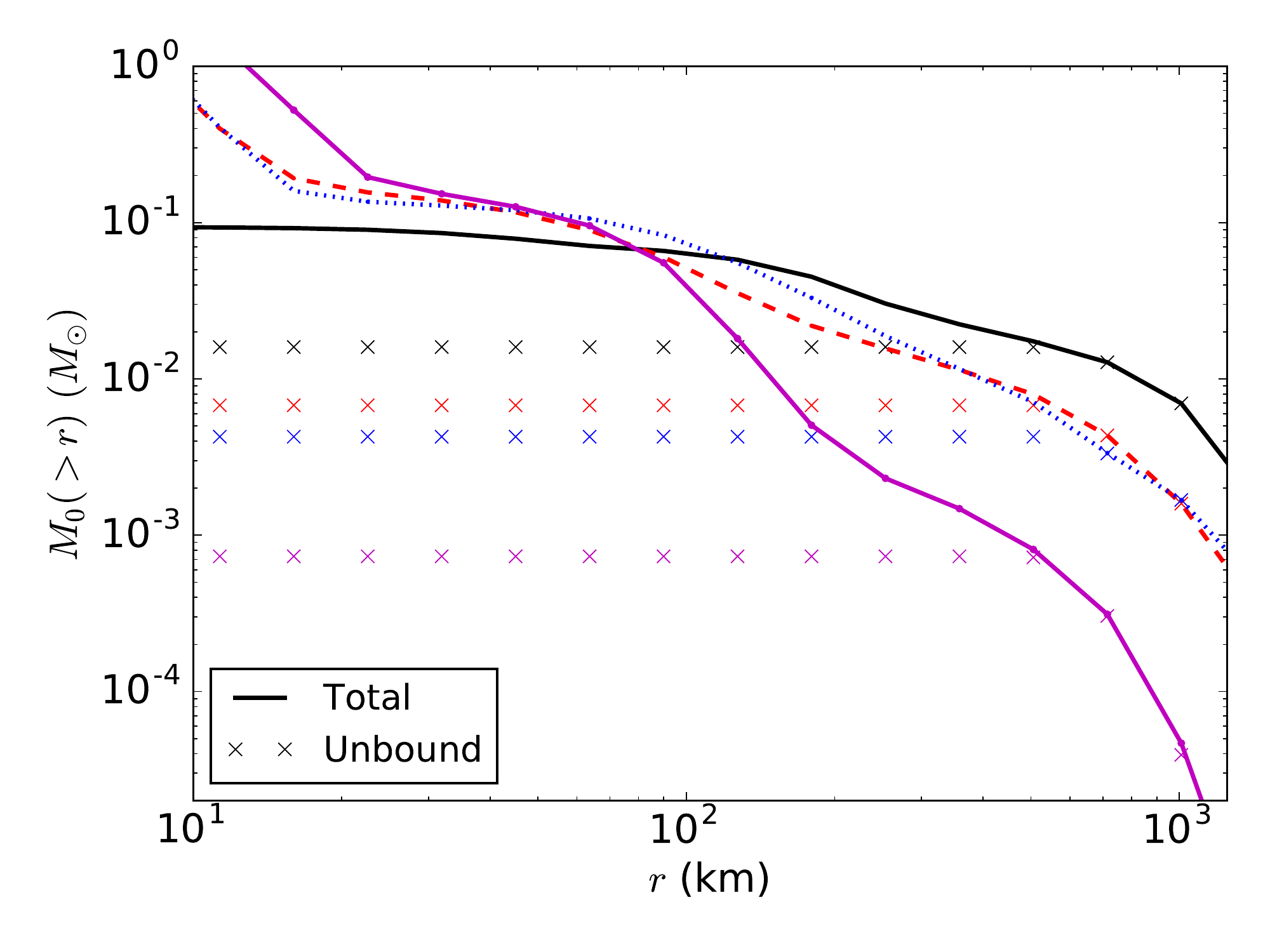}
	\includegraphics[width = 3.5in]{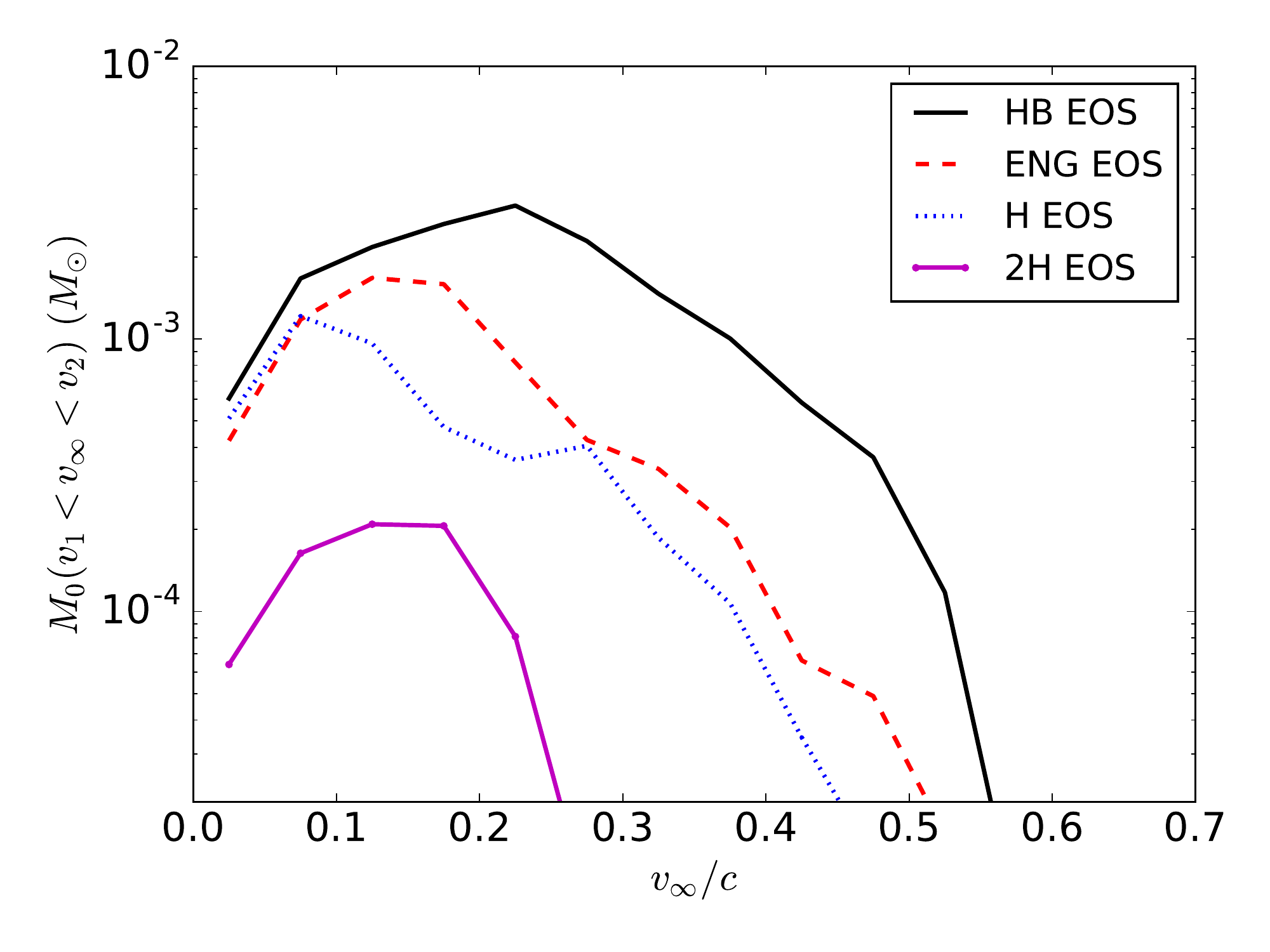}
\caption{ Left: The amount of rest mass outside a given coordinate radius from
  the center of mass for the post-merger remnant for mergers of
  nonspinning NSs with various EOSs.  The solid curves show the total
  amount, while the points show just the unbound rest mass.  Right:
  The amount of unbound rest mass binned by the velocity at
  infinity, with each bin $0.05c$ in size. The legend from the right panel applies to the left panel
  as well.  } \label{fig:ub_eos}
\end{center}
\end{figure*}

In this section, we examine the post-merger distribution of
matter. Comparing nonspinning cases with different EOSs, we see in
Fig.~\ref{fig:ub_eos} that stiffer EOSs give rise to more spread out
distributions of bound matter, but produce less unbound matter
compared to softer EOSs.  This trend in ejecta has been noted in
numerous studies of BNS mergers (see, e.g.,~\cite{2013PhRvD..87b4001H,Sekiguchi:2016bjd,2016MNRAS.tmp..894R},
and~\cite{Baiotti_Rezzolla_review2016} for a review), and can be
attributed to the fact that smaller radius NSs collide at higher
velocities, and thus produce more shock-heated dynamical ejecta.

\begin{figure*}
\begin{center}
	\includegraphics[width = 3.5in]{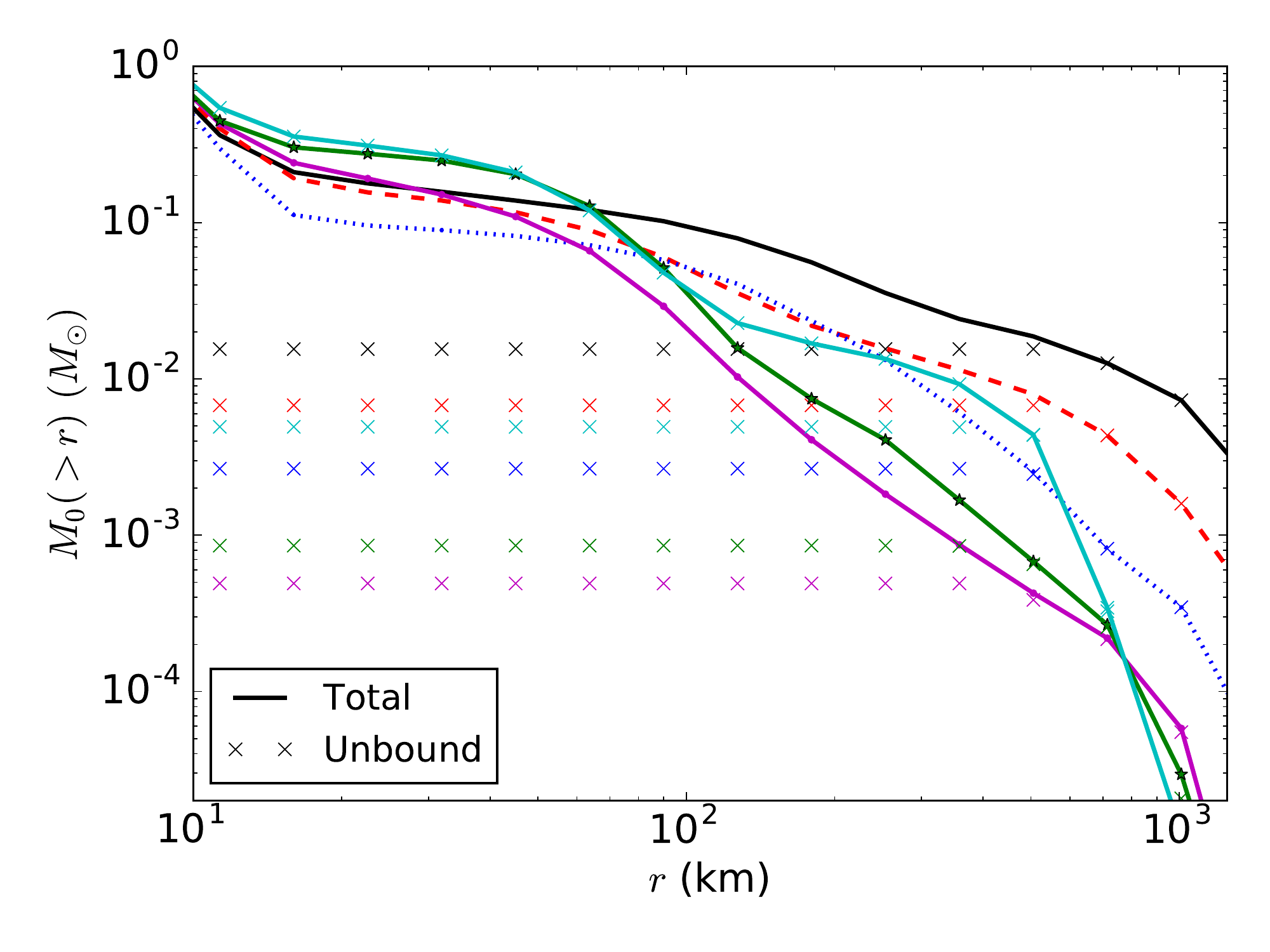}
        \includegraphics[width = 3.5in]{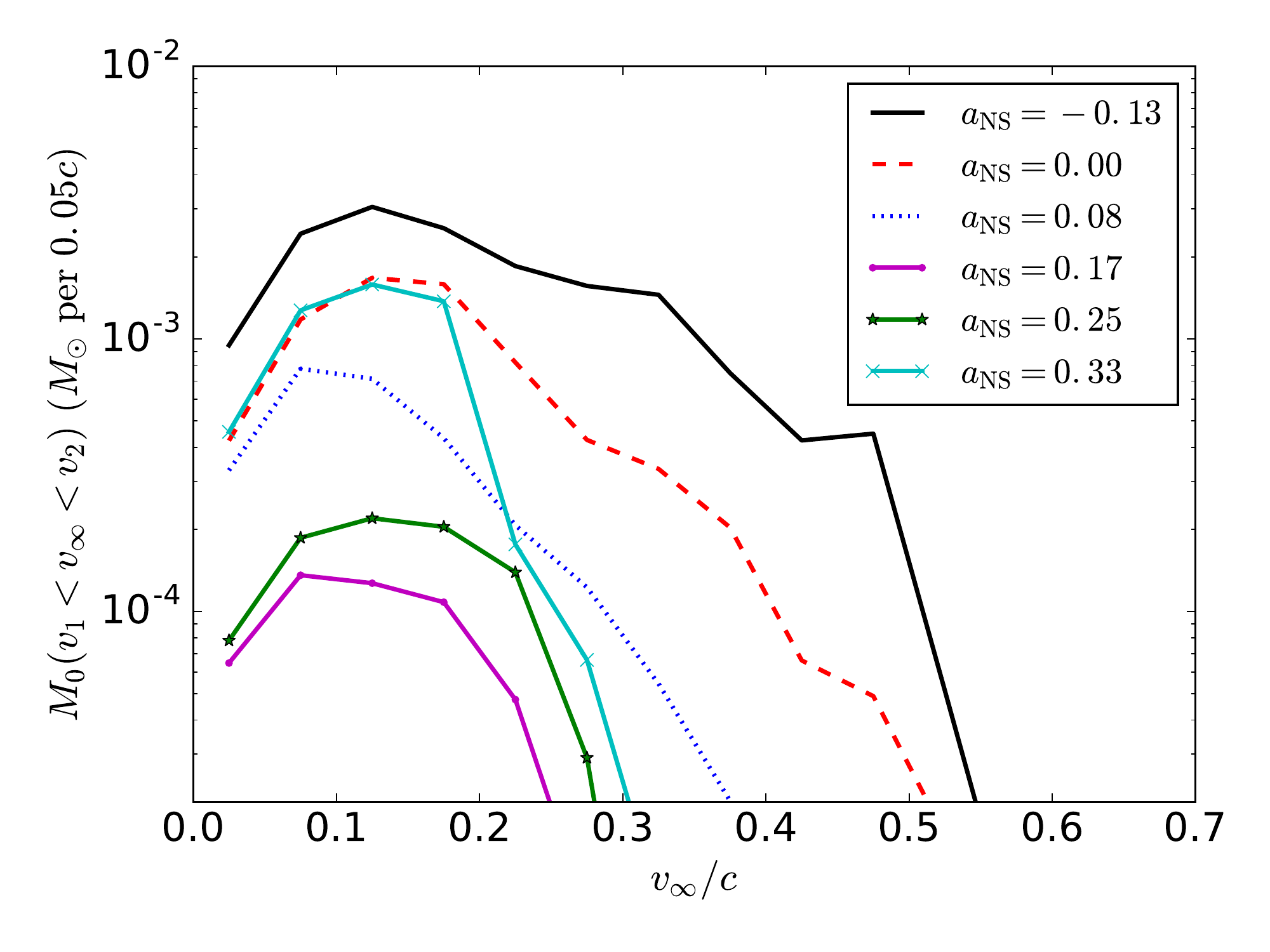}
\caption{ Same as Fig.~\ref{fig:ub_eos}, but for cases with the ENG
  EOS and various values of NS spin.  The legend from the right panel
  applies to the left panel as well.  } \label{fig:ub_eng}
\end{center}
\end{figure*}

The trend in unbound matter with NS spin is less clear cut.  As shown
in Fig.~\ref{fig:ub_eng}, the greatest amount of unbound material for
the ENG EOS ($\sim2\%$ $M_{\odot}$) is found for spins antialigned
with the orbital angular momentum. A similar result was also reported
in~\cite{Most:2019pac}. Our cases with increasing spin aligned with
the orbital angular momentum show decreasing amounts of unbound
material up to $a_{\rm NS}\approx 0.17$, at which point the trend
reverses, with the highest spin cases showing increased amounts of
unbound material up to $a_{\rm NS}=0.33$, which shows a similar amount
of unbound material to the nonspinning case. The fact that higher
negative spin results in more ejected matter is likely related with
the fact that in such cases the NSs plunge and collide at larger
velocities, ejecting more matter.  On the other hand, for sufficiently
rapidly rotating NSs (certainly as one approaches breakup spins) one
can also anticipate that it will be easier to tidally unbind the outer
NS matter.

\begin{figure*}
\begin{center}
	\includegraphics[width = 0.32\linewidth]{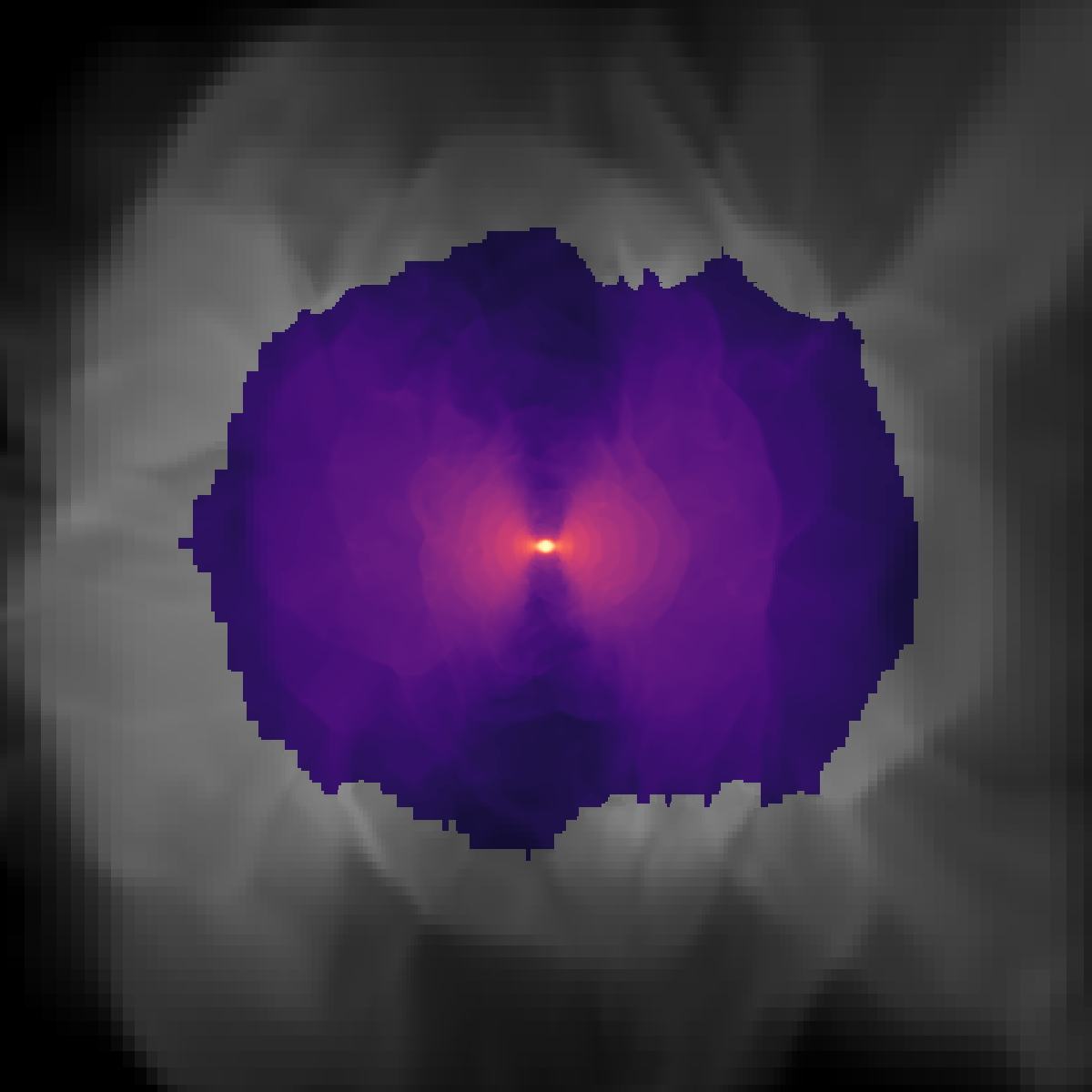}
	\includegraphics[width = 0.32\linewidth]{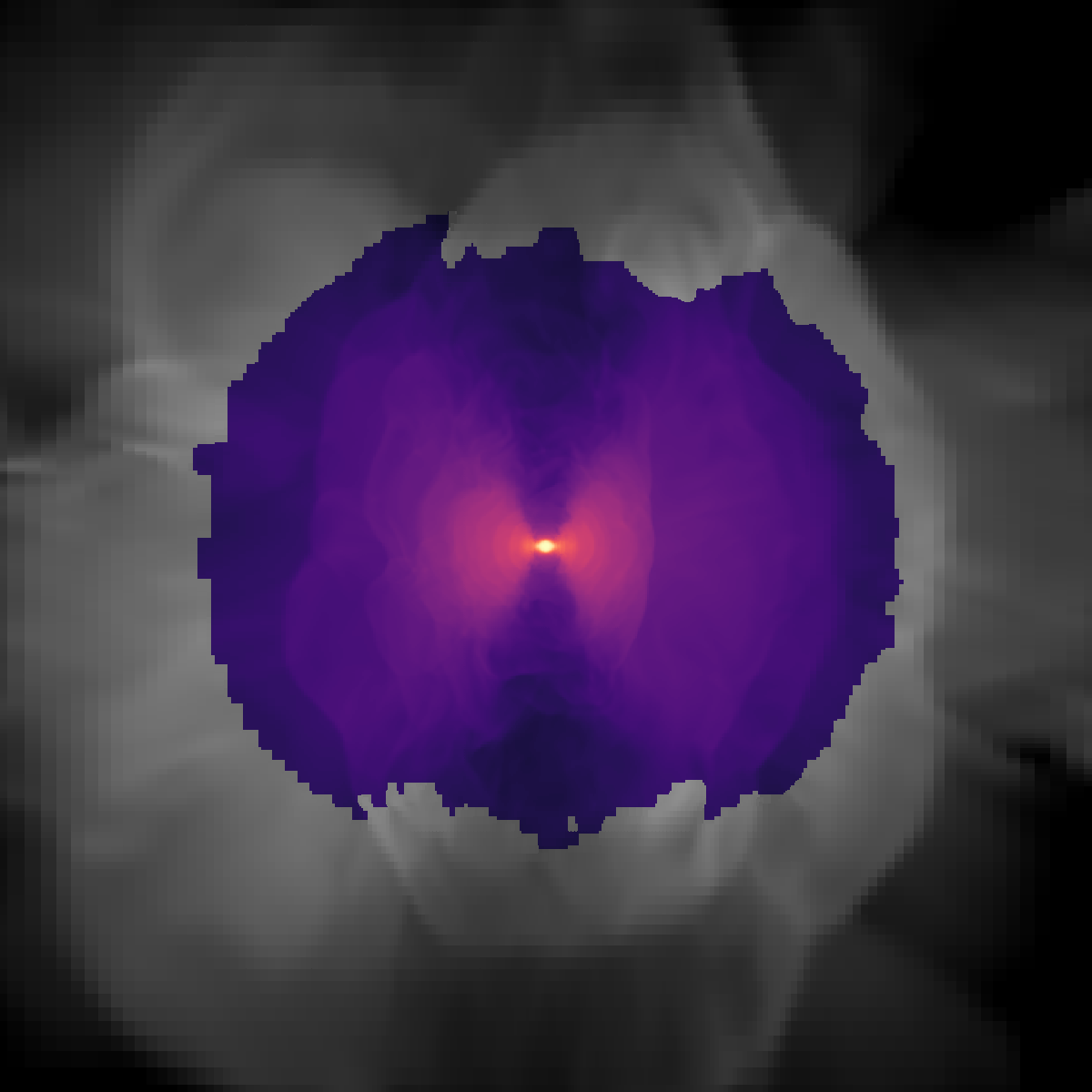}
	\includegraphics[width = 0.32\linewidth]{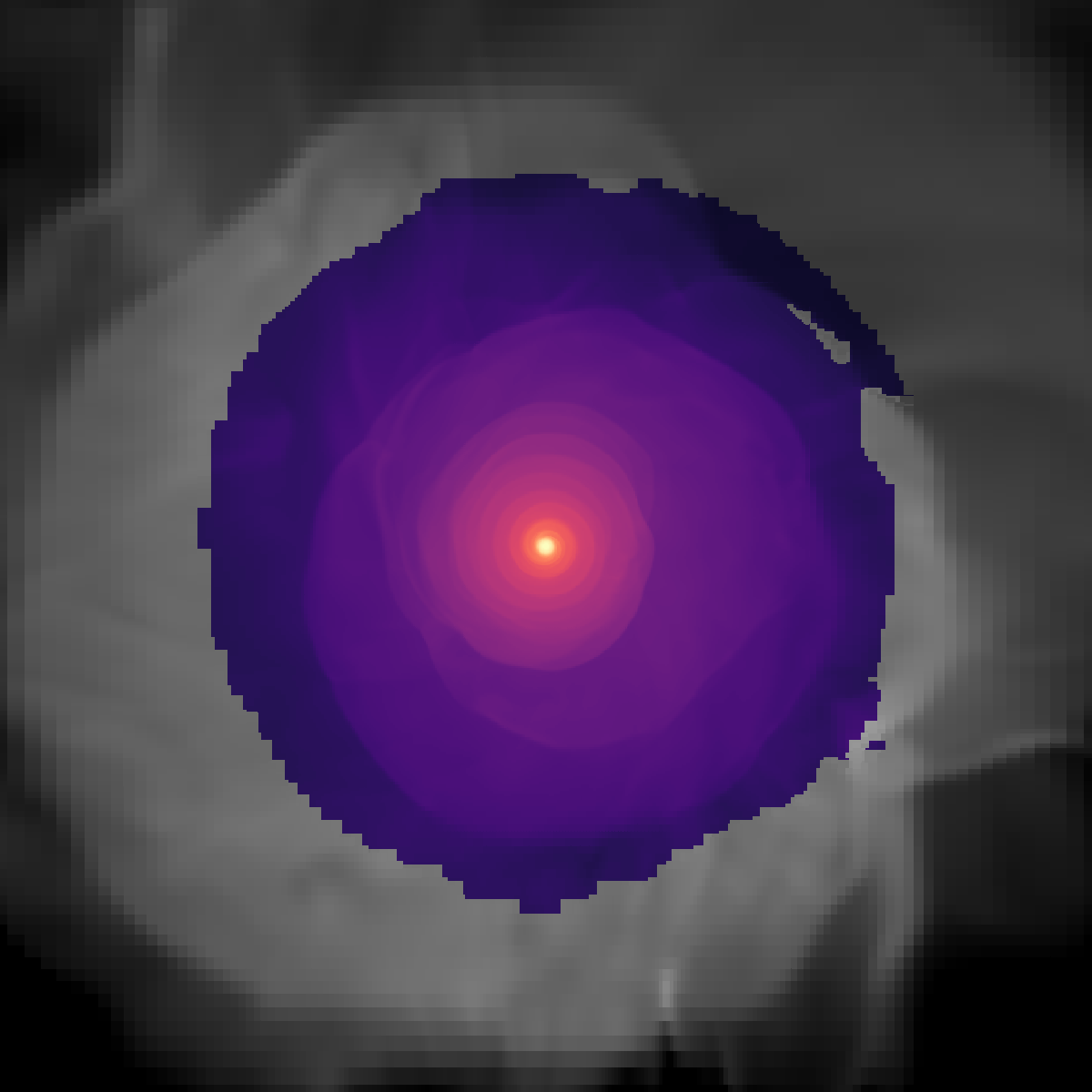}
	\includegraphics[width = 0.32\linewidth]{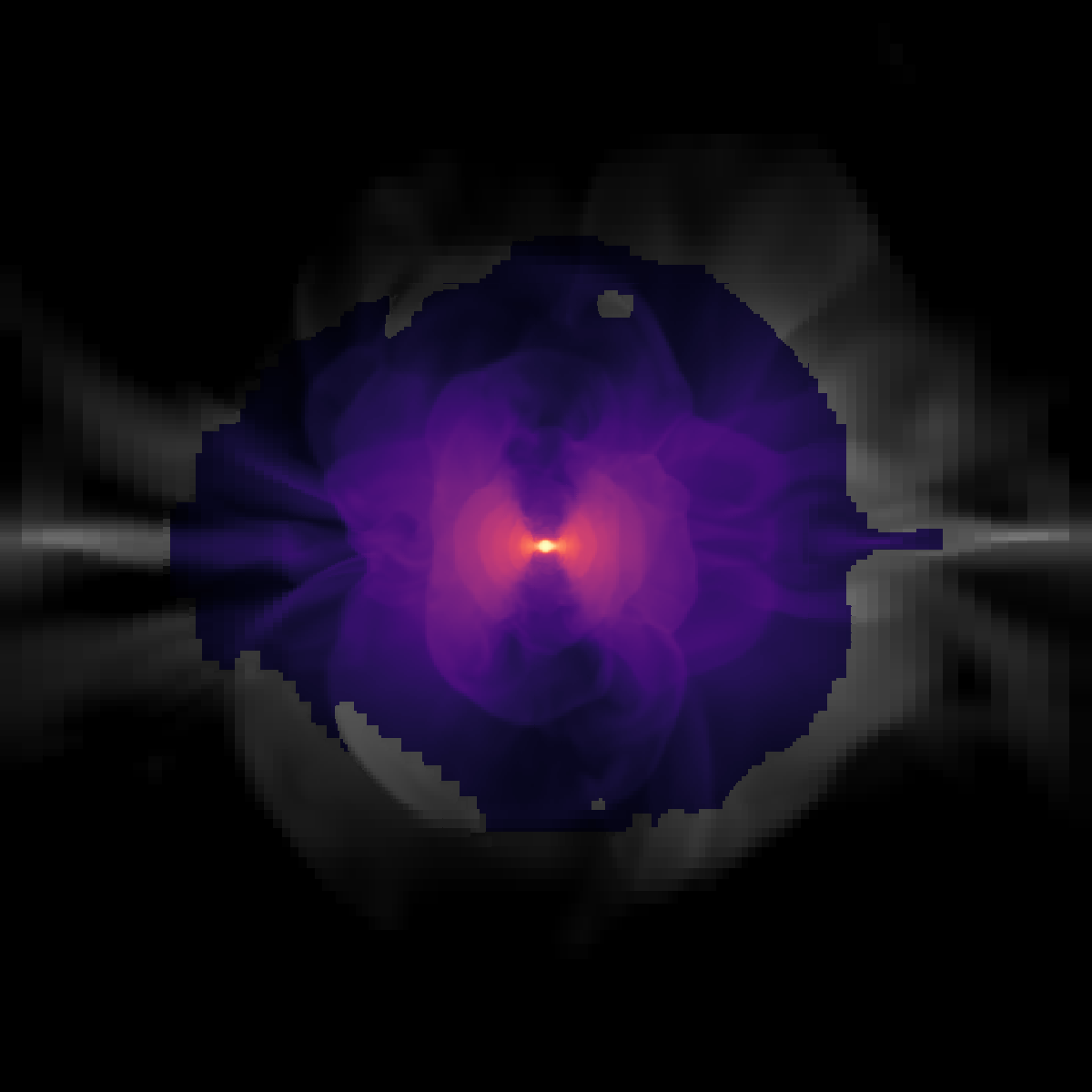}
	\includegraphics[width = 0.32\linewidth]{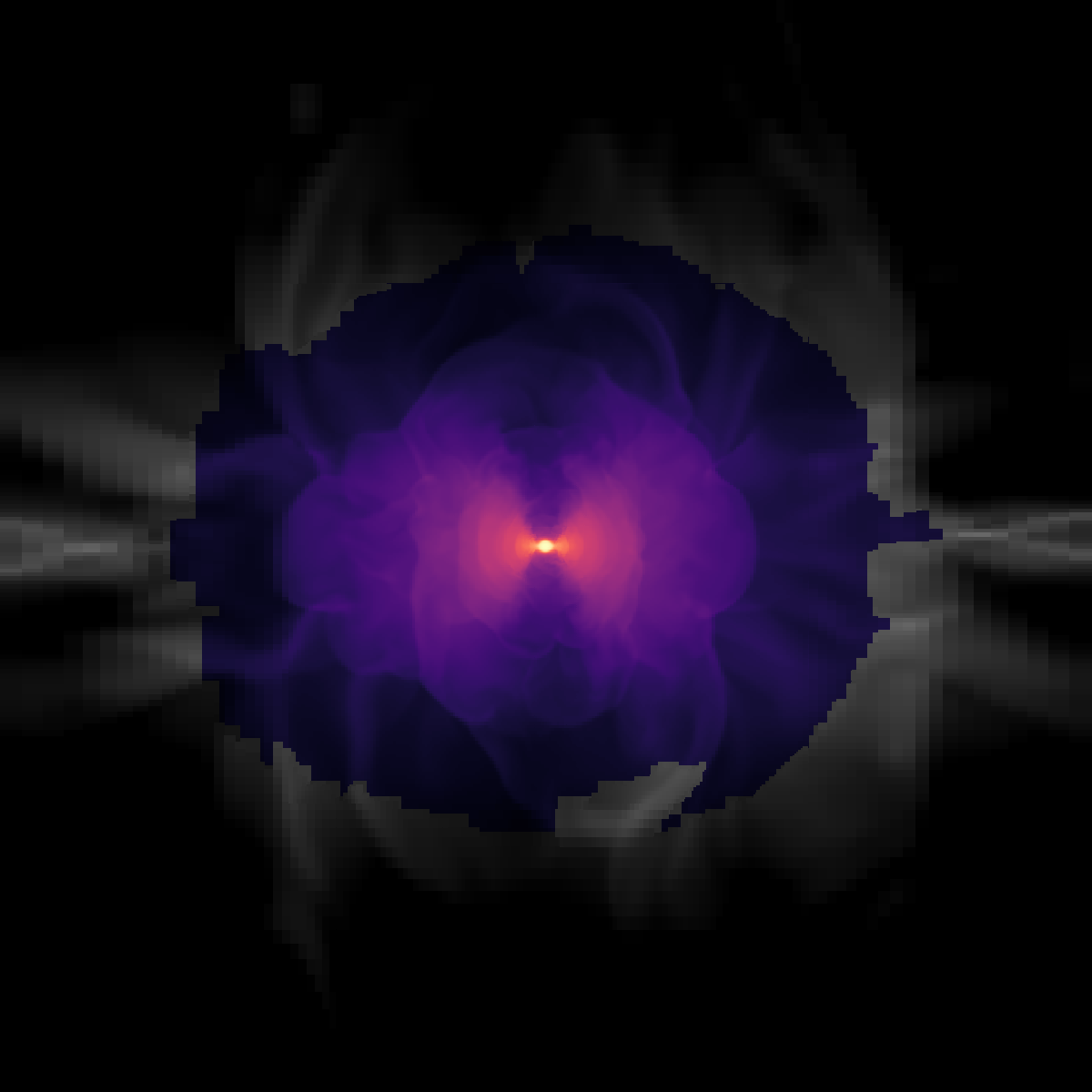}
	\includegraphics[width = 0.32\linewidth]{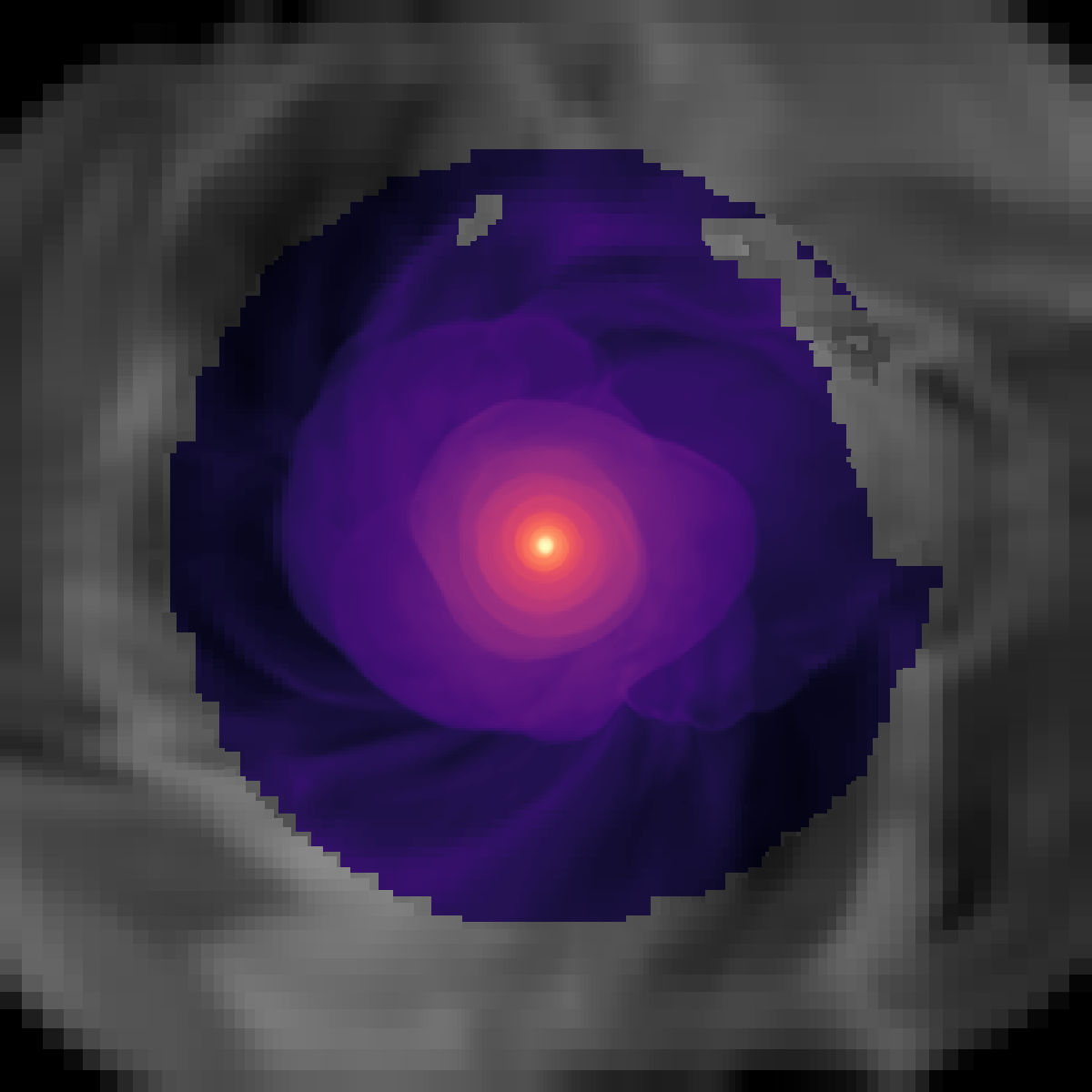}
	\includegraphics[width = 0.32\linewidth]{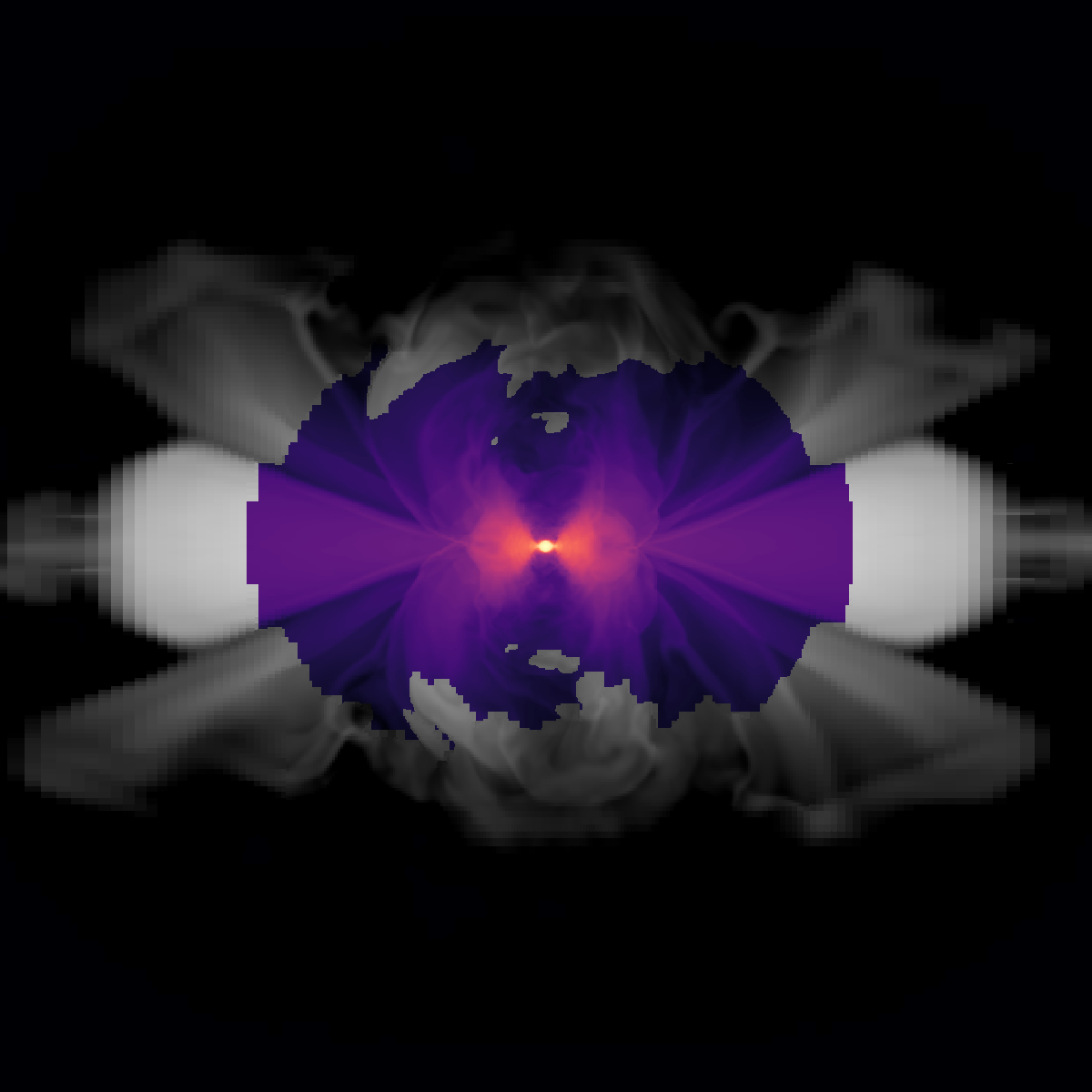}
	\includegraphics[width = 0.32\linewidth]{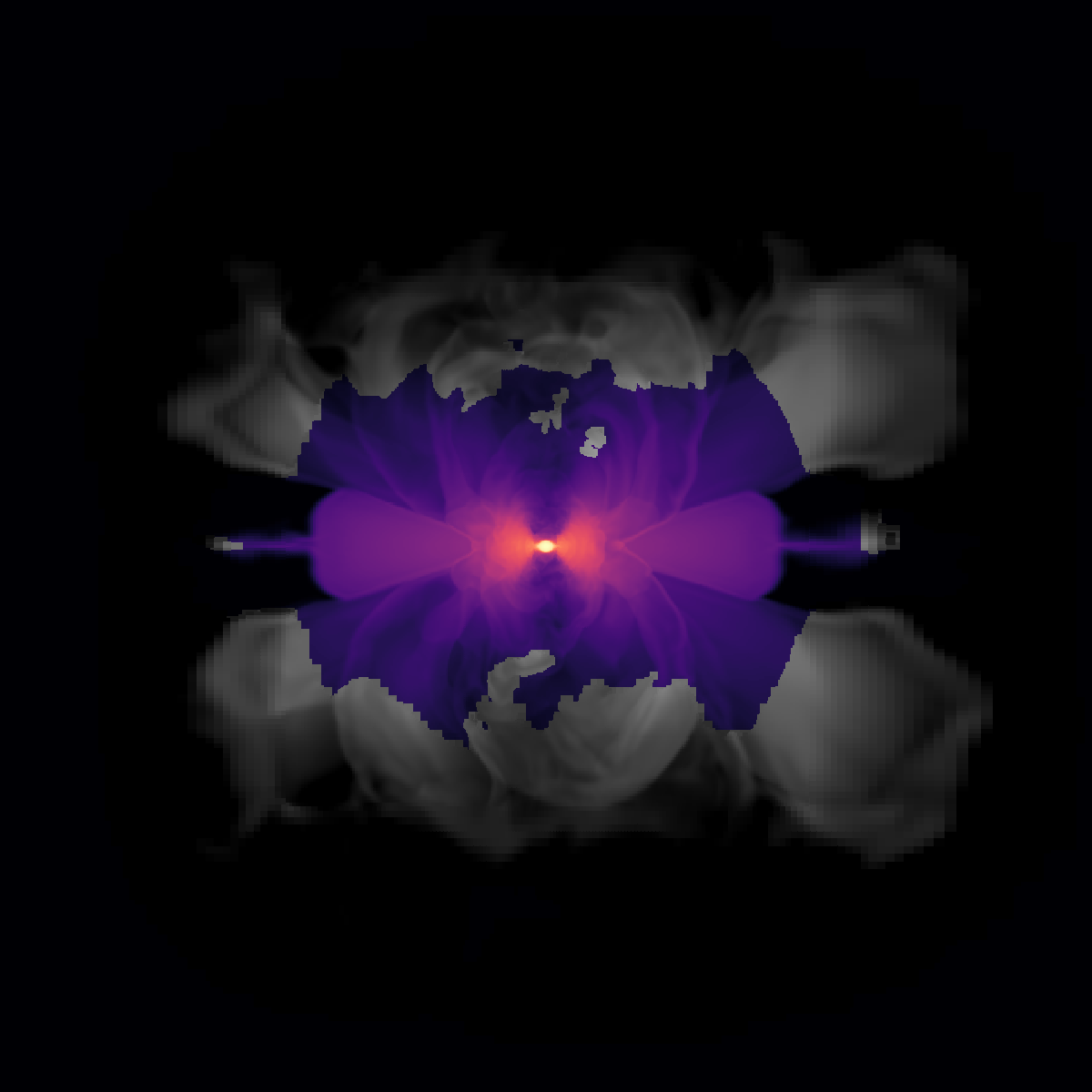}
	\includegraphics[width = 0.32\linewidth]{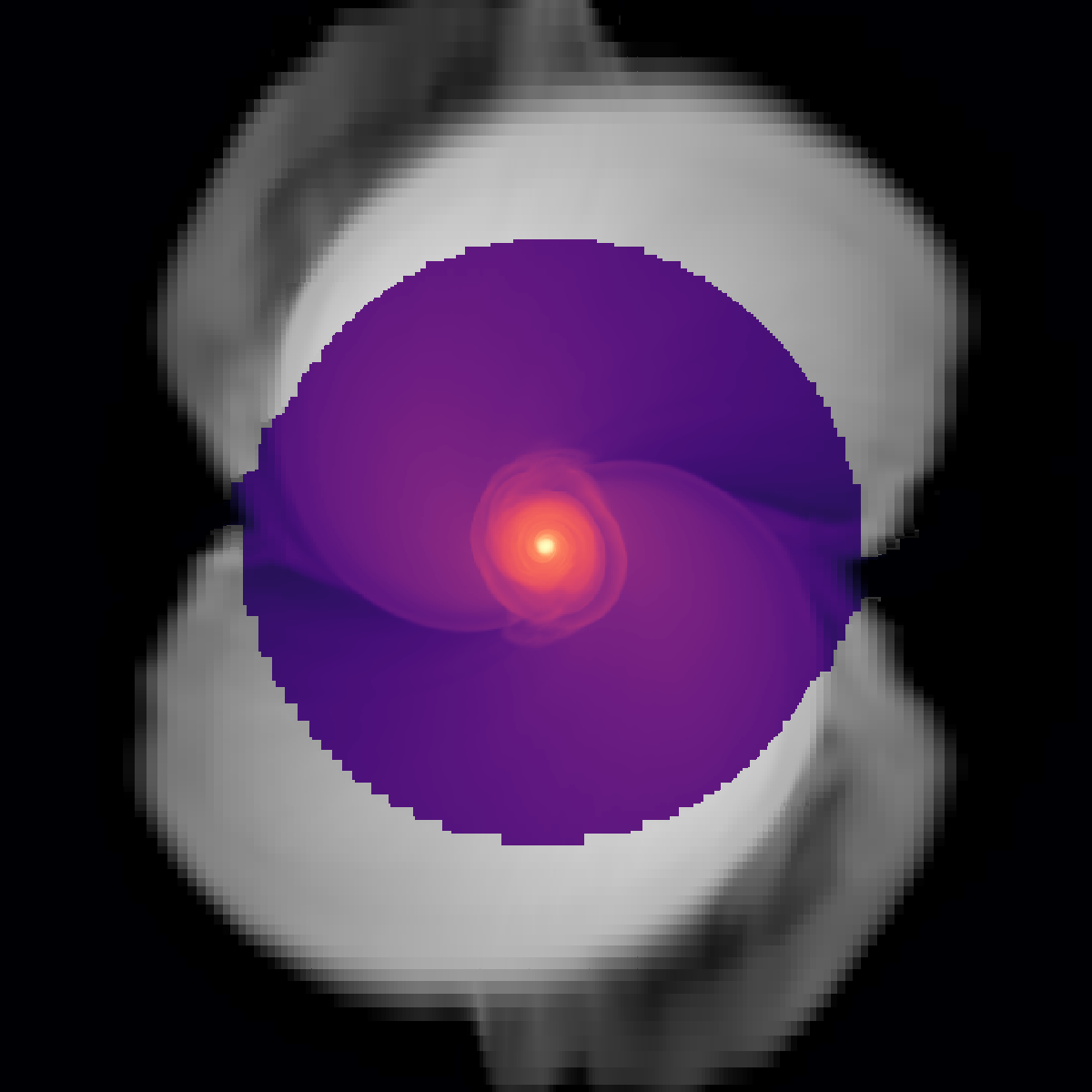}
    \caption{ The rest-mass density distribution ($D:=\rho_0u^t\sqrt{-g}$) of the bound (inner
   region; purple-yellow color scale) and unbound matter (outer
   region; black-white color scale) $\sim 10$ ms after merger. The
   columns show (left to right) slices in the $x=0$, $y=0$, and $z=0$
   (equatorial) plane. The rows (top to bottom) show cases with
   $a_{\rm NS}=0$, 0.17 (CO), and $0.33$. The plots show
   roughly $1800$ km in each linear dimension.
 } \label{fig:matter_pic}
\end{center}
\end{figure*}

In addition to affecting the total amount of mass of the ejecta, and
the ejecta mass distribution with velocity, spin substantially impacts
the angular distribution of dynamical ejecta. In
Fig.~\ref{fig:matter_pic}, we show snapshots of rest-mass density on
the $x=0$ (left column), $y=0$ (middle column), and $z=0$ (equatorial;
right column) planes with the zero pre-merger spin on the top row,
$a_{\rm NS}=0.17$ in the middle row, and $a_{\rm NS}=0.33$ on the
bottom row. The figure demonstrates that as the aligned spin
increases, the unbound matter becomes more concentrated near the
orbital plane, consistent with being due to tidal effects. By
contrast, for smaller spins, the ejecta are more isotropically
distributed.

A similar result showing that antialigned spin increased the amount
of dynamical ejecta, while aligned spins up to $\sim 0.3$ decreased
the amount ejecta and caused it to be more concentrated toward the
equatorial plane, was also found recently in Ref.~\cite{Most:2019pac}. In
contrast to what we find here, in that case no enhancement in ejecta
was found for very high spins.  However,~\cite{Most:2019pac} used
finite-temperature EOSs (and slightly lower maximum spin values), and
so is not directly comparable to this work. Moreover,
Ref.~\cite{Kastaun:2016elu} did not find the greatest amount of ejecta
for the antialigned spin case. However, that work not only used
different EOSs, but also constraint-violating and
nonequilibrium initial data, so that work is also not directly comparable
to the present study.

Apart from larger dynamical ejecta masses, we also find that NS spins
that are antialigned with the orbital angular momentum lead to a
substantial amount of matter moving outward at higher velocities (see
right panel of Fig.~\ref{fig:ub_eng}). By contrast, larger aligned
spin, decreases the width of the velocity distribution up to the spin
value of 0.17, above which the distribution width increases again. The
fact that in the antialigned case there is a tail of the ejecta
moving at high velocities, implies that the radio emission from
antialigned spin BNS mergers can be significantly brighter compared
to irrotational mergers. In particular, Fig.~\ref{fig:ub_eng} suggests
that in the antialigned case about $10^{-3} M_\odot$ of matter moves
at speed greater than 0.3c, which implies that this tail alone has a
total kinetic energy of about $10^{51}$ erg. Therefore, based on
Eqs.~\eqref{EjectaISMtime} and~\eqref{Fnu}, mergers with antialigned
spin are likely to have radio emission from the interaction with the
interstellar medium which is significantly brighter with shorter rise
times than irrotational and aligned-spin cases.

\label{sec:matter}

This trend of the ejecta mass distribution and ejecta
velocities with spin would cause the anticipated kilonova component that is
powered by dynamical ejecta to be significantly brighter in the
antialigned case. In addition, the estimated kilonova in the
irrotational case is brighter than all cases with aligned spin that we
study. The latter results holds despite the fact that the amount of
dynamically ejected mass in the spin 0.33 and irrotational cases is
approximately the same.  We detail the properties of the unbound
material in Table~\ref{matter_table}, along with an estimate of how
these may translate into differences in the ejecta-powered kilonovae
by use of Eq.~\eqref{Lkilonovae}. As can be seen in the table, keeping
all the other parameters fixed, NS spin can make an order of magnitude
difference in the ejecta properties, including the kilonova rise
time. However, its influence is degenerate with other parameters, such
as the NS EOS, and most likely also depends on the mass ratio, which we do
not treat here.
We also note that there is almost a factor of 2 difference in the amount
of unbound material for the two different $a_{\rm NS}=0.17$ cases, though
in both cases the total mass is small compared to the other cases.

In recent years, a number of
studies~\cite{Fernandez:2013tya,Metzger:2014ila,Perego:2014fma,Just2015MNRAS.448..541J,Wu:2016pnw,Lippuner:2017bfm,Siegel:2017nub}
have suggested that a large fraction of the mass of the disk that
forms around the BNS merger remnant becomes unbound due to the effects 
of viscosity, neutrinos, and/or magnetic fields. These disk winds
contribute to the kilonova signature, and appear necessary to explain
GW170817. Therefore, the amount of mass that forms a disk around the
merger remnant crucially determines the kilonova properties,
as well as, presumably, affecting any accompanying gamma-ray burst. In
Table~\ref{matter_table}, we list the amount of matter in the disk
around the merger remnant. As with the dynamical ejecta, the trend of
the disk mass with spin is not monotonic. However, the highest
aligned-spin case has the most massive accretion disk. This result is
in agreement with the findings of~\cite{Ruiz:2019ezy} where a simpler
$\Gamma$-law EOS with $\Gamma=2$ was adopted, but
magnetic fields were treated. Here we find that, compared to the
irrotational case, dimensionless spins of $\sim0.3$ can double the amount of mass in
the disk around the merger remnant. Therefore, high aligned spin makes
it easier to form large accretion disks that can contribute to the blue
component of kilonovae.

\begin{table*}[t]
\caption{\label{matter_table} The properties of the bound and unbound
  NS matter from various cases. The columns list the EOS, 
  the dimensionless NS spins $a_{\rm NS}$ (which are the same
  for both stars), the bound rest mass $M_{0,\rm disk}$ with $r>32$ km in units of
  $M_\odot/100$ and the radius $R_{\rm disk}$ inside which $90\%$ of this mass is
  contained in units of $10^5$ m, the unbound rest mass $M_0$ in units
  of $M_\odot/100$, the average asymptotic velocity weighted by
  rest mass $\langle v_{\infty}\rangle$, the total kinetic energy in
  units of $10^{50}$ erg, the anticipated kilonovae rise time in days,
  and the associated luminosity in units of $10^{41}$ erg/s, which we
  compute via Eqs.~\eqref{tkilonovae} and~\eqref{Lkilonovae}.  
  See Appendix~\ref{app:convergence} for details regarding measuring these quantities.
    }
\centering
\begin{tabular}{cccccccccccc}
\hline\hline
EOS &
$a_{\rm NS}$ &
Spin state &
$M_{0,\rm disk}$ & 
$R_{\rm disk}$ & 
$M_{0,\rm u}$ & 
$\langle v_{\infty}\rangle$ &
$E_{{\rm kin},50}$ &
$t_{\rm peak}$ &
$L_{41}$
\\
\hline
 ENG   &  $-0.13$   &   SP   &  14   & 3.1   &  1.55   &   0.20   &   7.88 &   0.37   &  1.98  \\
 ENG   &  0.00   &   IR   &  13   & 2.0   &  0.68   &   0.17   &   2.36 &   0.23   &  1.20  \\
 ENG   &  0.08   &   SP   &  9   & 2.8   &  0.27   &   0.13   &   0.55 &   0.12   &  0.66  \\
 ENG   &  0.17   &   SP   &  15   & 1.2   &  0.05   &   0.12   &   0.08 &   0.05   &  0.28  \\
 ENG   &  0.17   &   CO   &  17   & 1.3   &  0.09   &   0.12   &   0.15 &   0.07   &  0.37  \\
 ENG   &  0.25   &   SP   &  25   & 1.2   &  0.09   &   0.14   &   0.19 &   0.07   &  0.39  \\
 ENG   &  0.33   &   SP   &  26   & 1.1   &  0.49   &   0.12   &   0.81 &   0.17   &  0.88  \\
 2H   &  0.00   &   IR   &  15   & 1.4   &  0.07   &   0.13   &   0.14 &   0.07   &  0.35  \\
 H   &  0.00   &   IR   &  12   & 2.8   &  0.43   &   0.15   &   1.23 &   0.17   &  0.89  \\
 HB   &  0.00   &   IR   &  7   & 3.5   &  1.60   &   0.22   &   9.53 &   0.40   &  2.13  \\

\hline\hline 
\end{tabular}
\end{table*}

\section{Conclusions}
\label{conclusions}
In this work, we presented results from fully relativistic hydrodynamic simulations
of quasiequilibrium BNSs in quasicircular orbits with {\it
  spinning} components. We considered configurations of equal mass and
equal spin, with initial dimensionless NS spins in the range~$a_{\rm
  NS} \in [-0.13,\,0.33]$. We modeled the matter as a perfect fluid,
with different piecewise polytropic representations for the equation
of state, covering a range of compactness for a 1.4 $M_\odot$ NS from
$\sim $0.136 to 0.178. We focused on quantifying 
the effects of pre-merger NS spin, and
neglected the effect of magnetic fields, neutrinos, and realistic
nuclear microphysics in this first study.

We find that the basic features of the GW spectrum following merger
are broadly consistent with what has been found in previous works
(see~\cite{Paschalidis:2016vmz} for a review). More specifically, we
find that the post-merger peak GW frequency is only weakly influenced
by the NS spin (by about 100--200 Hz) consistent with
Refs.~\cite{Bernuzzi:2013rza,PhysRevD.95.044045,bauswein2015exploring}.
We find that aligned (antialigned) spin cases give slightly lower
(higher) values of the post-merger peak GW frequency when compared to
the irrotational case. At higher frequencies (next to the peak GW
frequency) antialigned configurations enhance the GW power with
respect to aligned ones. In turn, these results imply that there is
some degree of degeneracy between pre-merger spin and the nuclear EOS,
and without some independent measurement of spin this should
contribute to the systematic error when inferring the EOS from the
post-merger peak GW frequency.

We find that the one-arm instability can also operate in the remnant
of quasicircular mergers with spin. Spin affects the GW frequency
associated with the one-arm mode in the same way it affects the peak
post-merger GW frequency, i.e., aligned spin shifts the one-arm mode
frequency to lower values. Our studies do not show a significant
correlation between the energy that goes into the one-arm mode and the
pre-merger spin. We find that the strongest one-arm mode develops for
an initially corotating case that we consider. The GW signal from this
one-arm mode could be detectable by third-generation GW detectors, and
the one-arm mode GW frequency can be used to infer properties of the
nuclear EOS.

Our simulations demonstrate that spin has a substantial impact on the
mass, velocity, and angular distribution of dynamical ejecta, that would
likely be reflected in the red kilonova signatures from such events. 
Our results also indicate that spins
antialigned with the orbital angular momentum result in more massive
dynamical ejecta, with a considerable amount of matter traveling at
speeds near 0.5c (see also~\cite{Most:2019pac}). Antialigned spin mergers generate brighter red
kilonovae than aligned-spin cases, because the latter have suppressed
dynamical ejecta masses. However, we find that as the aligned spin
increases past a certain value, the amount of dynamical ejecta
increases again (but the velocity distribution of ejecta masses 
is narrower than the irrotational and antialigned cases). This implies
that the expected red kilonovae should again become brighter as the
spin increases---consistent with the fact that as the spin frequency
increases, the star becomes less bound, and hence it becomes easier to
dynamically eject more mass. For higher spins, we find that  
the dynamical ejecta are more concentrated around the orbital plane.

The blue kilonova expected from unbinding part of the remnant disk is also
affected by the initial NS spin insofar as the latter affects the disk mass
and/or the lifetime of the remnant. Recent
work~\cite{Kasen:2014toa,Fernandez:2015use,Siegel:2017nub} has shown that a
substantial amount of mass outside the remnant becomes unbound due to
viscous/magnetic/neutrino processes. For aligned dimensionless spins of
$\sim0.2$--$0.3$, the merger remnants have larger disks than the lower-spin
cases, and the massive NS remnants from aligned spin BNS mergers likely survive
for longer times than those from nonspinning mergers due to the extra
centrifugal support provided by the additional total angular momentum. The
longer the massive NS remnant survives, the larger the unbound disk mass due to
strong neutrino irradiation from the hot remnant can be.  Also, assuming that
the fraction of the disk mass that becomes unbound is approximately independent
of the mass of the disk that initially forms around the remnant, we anticipate
that the blue kilonovae should be brighter in NS mergers with higher aligned
initial spins.  However, this conjecture should be carefully studied with
long-term viscous/MHD studies that treat neutrino heating. Studies which
attempt to place a constraint on the binary tidal deformability $\tilde
\Lambda$ based on how much mass goes into a disk structure have so far not
included the effects of spin.  For example, the work of~\cite{Radice:2017lry}
suggests that for equal mass binaries, $\tilde \Lambda\gtrsim 400$ is necessary
to explain the kilonova counterpart to GW170817, while this value is lowered in
the unequal mass case~\cite{Kiuchi:2019lls}. These studies may need to be
revisited to include the effects of initial NS spin, and the bounds on $\tilde
\Lambda$ are likely to become less restrictive for spinning BNSs. We point out
that in our work we do not study asymmetric binaries, e.g., unequal masses or
unequal spins, nor do we treat magnetic fields, neutrinos, or other such
microphysics effects, all of which could change some of our results.  We intend
to address these points in future work.

Finally, we compared two simulations that have the same initial properties,
i.e., the same total mass, orbital angular frequency, and circulation, 
but one corresponds to corotating initial data and the other to the
corotating counterpart constructed with the constant rotational-velocity
formulation~\cite{Tichy:2011gw,Tichy:2012rp}. The two binary systems are broadly
equivalent, though we find some differences in the merger time
and post-merger matter distribution. This could be due to the fact that their 
identification through the concept of equatorial circulation is not exact
\cite{Tsokaros:2018dqs}. In turn the outer layers of the stars may have slightly
different properties, which would lead to observed differences in dynamical ejecta 
masses, etc.


\acknowledgments
W.E.E. acknowledges support from an NSERC Discovery grant.
V.P. acknowledges support from NSF Grant
PHY-1912619. F.P. acknowledges support from NSF Grant PHY-1607449, the
Simons Foundation, and the Canadian Institute For Advanced Research
(CIFAR).  A.T. acknowledges support from NSF Grant PHY-1662211, and
NASA Grant 80NSSC17K0070. V.P. and F.P. would like to thank KITP for
hospitality, where part of this work was completed. V.P. would also
like to thank B. Metzger for useful discussions. This research was
supported in part by Perimeter Institute for Theoretical
Physics. Research at Perimeter Institute is supported by the
Government of Canada through the Department of Innovation, Science and
Economic Development Canada and by the Province of Ontario through the
Ministry of Economic Development, Job Creation and Trade. Research at
KITP is supported in part by the National Science Foundation under
Grant No. NSF PHY-1748958. Computational resources were provided by
XSEDE under Grant TG-PHY100053 and the Perseus cluster at Princeton
University.

\appendix

\section{Corotating versus spinning initial data}
\label{sec:covsspi}
Here we compare the two cases we consider that have the same
quasilocal value for the dimensionless NS spin, namely the ENG
$a_{\rm NS}=0.17$ ``SP'' and ``CO'' cases listed in
Table~\ref{table:NSNS_ID}. These two types of initial configurations
correspond to different formalisms, but their global properties are
essentially the same.

\begin{figure}
\begin{center}
	\includegraphics[width = 3.5in]{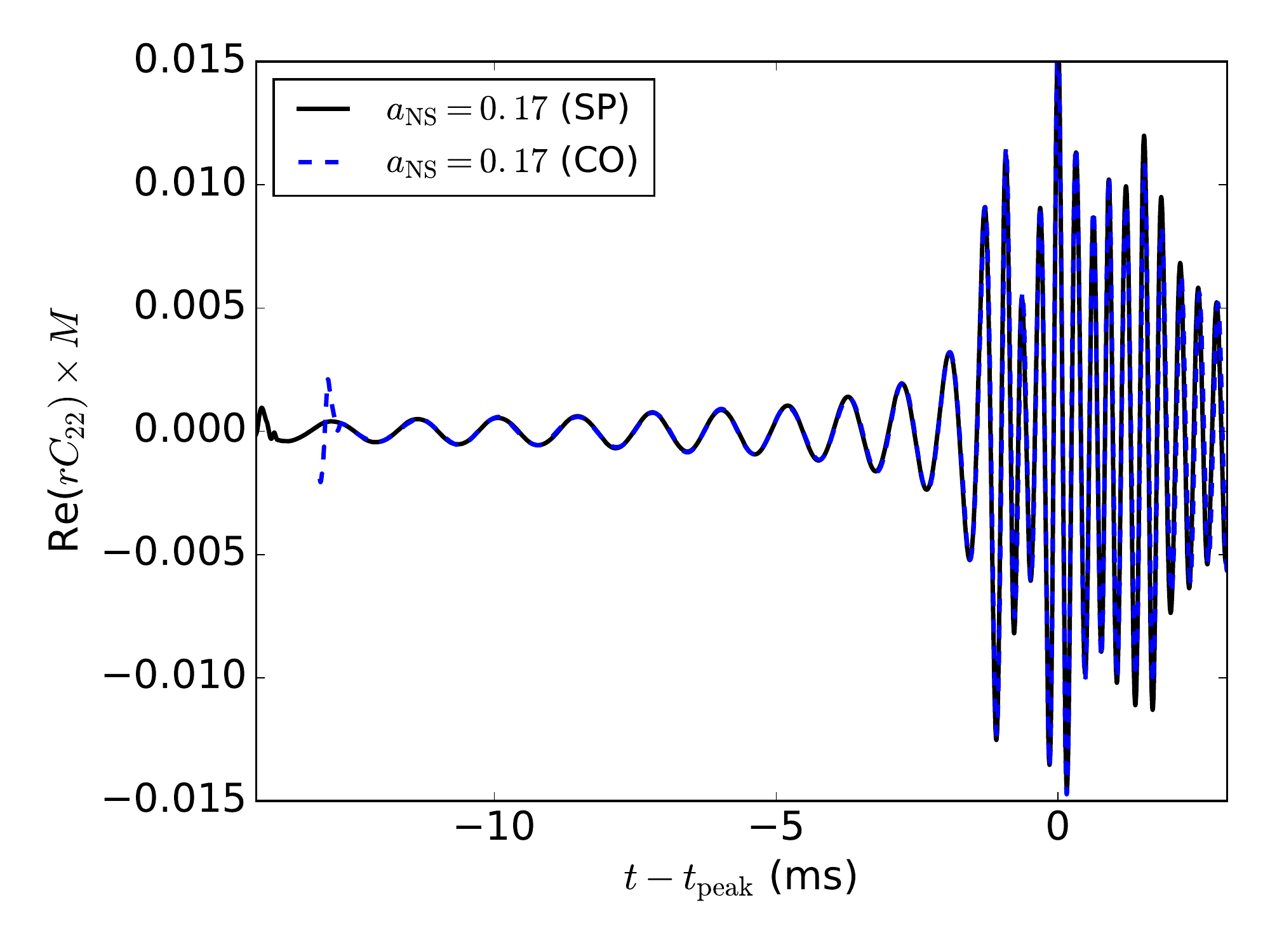}
	\includegraphics[width = 3.5in]{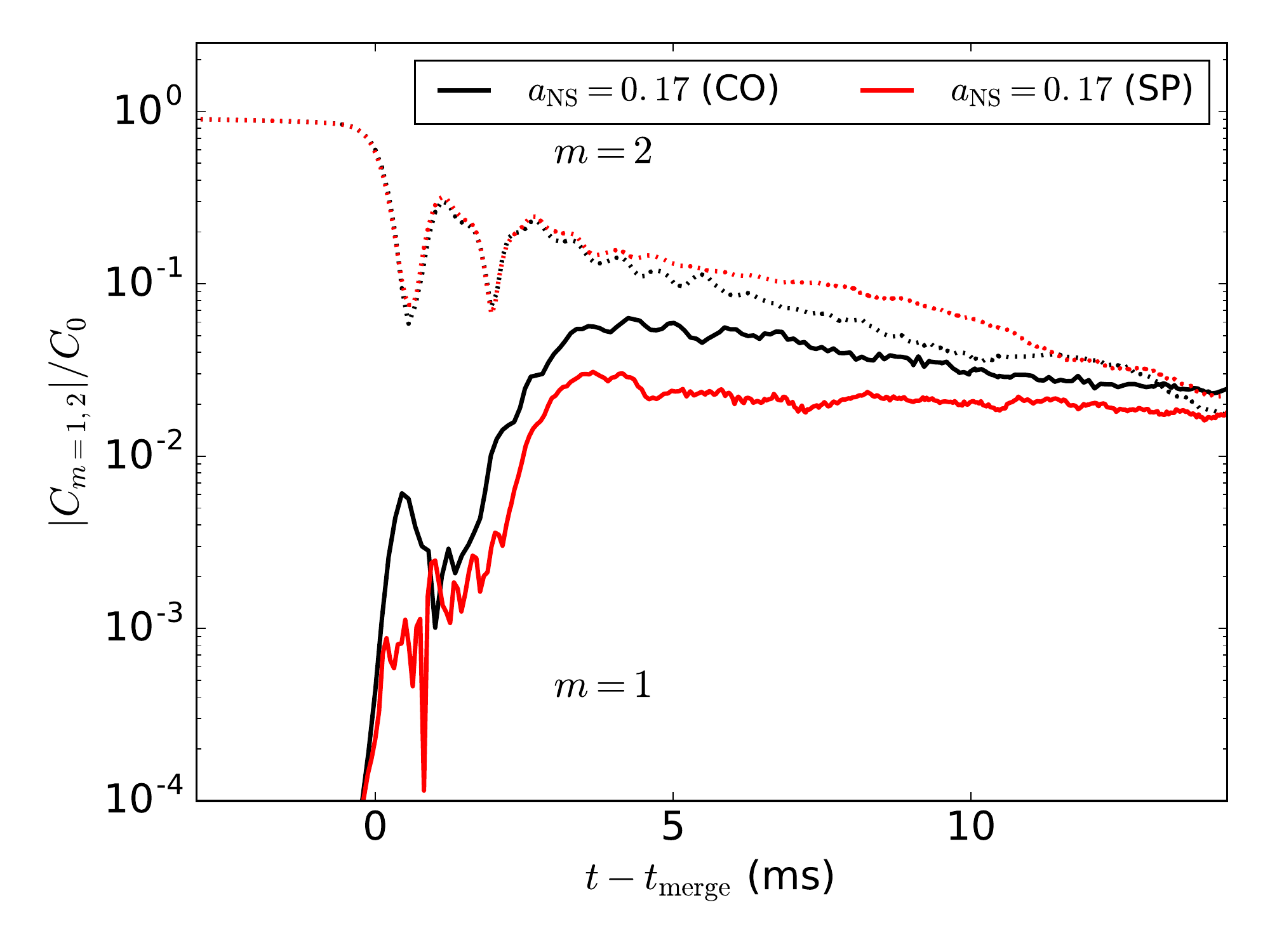}
\caption{
    Top: The GW signal ($\ell=m=2$ component of
    $\psi_4$ multiplied by the extraction radius $r$) from mergers with the ENG EOS and $a_{\rm NS}=0.17$ 
	but different spin formalisms. The
	curves have been aligned in time and phase at peak. 
    In particular, the CO case has been shifted ahead by 1.1 ms relative
    to the SP case.
    Bottom: The magnitude of the $m=1$ and $m=2$ azimuthal density modes as a function
    of time for the same two cases. 
 } \label{fig:co_sp}
\end{center}
\end{figure}

Despite the global similarities in the two types of initial data, we
find that their evolution has some differences. To begin with, the SP
case undergoes an additional $\sim1/3$ of orbit before merging, though
accounting for this difference in the time (and phase) of the merger,
the GW signals otherwise line up well as shown in the top panel of
Fig~\ref{fig:co_sp}.  The disk mass that forms in the two cases
differs by about $10\%$, but the total dynamical ejecta in the CO case
is about 2 times as massive as in the SP case (though in either case
it is small). The average ejecta velocity is the same in the two
cases, but the kinetic energy is a factor of 2 different.  There is
also some difference between the amplitudes of the azimuthal density
modes produced post-merger, with the SP case exhibiting slightly
larger $m=2$ and slightly smaller $m=1$ modes relative to the CO case,
as shown in bottom panel of Fig.~\ref{fig:co_sp}. However,
higher-resolution simulations are probably required to determine how
much of this difference is numerical.

The small differences between the two types of initial data suggest
that the ``corotation limit'' of the constant rotational-velocity
formulation agrees well with the corresponding corotating
formulation. The bulk of the matter behaves the same in the two
formulations, but it is likely that the outer NS layers in the
constant rotational-velocity formulation have slightly different
properties, probably as a result of their identification through their 
circulation. In turn this would explain the difference in dynamical 
ejecta that we find.

\section{Numerical errors and convergence study}
\label{app:convergence}

\begin{figure*}
\begin{center}
	\includegraphics[width = 3.5in]{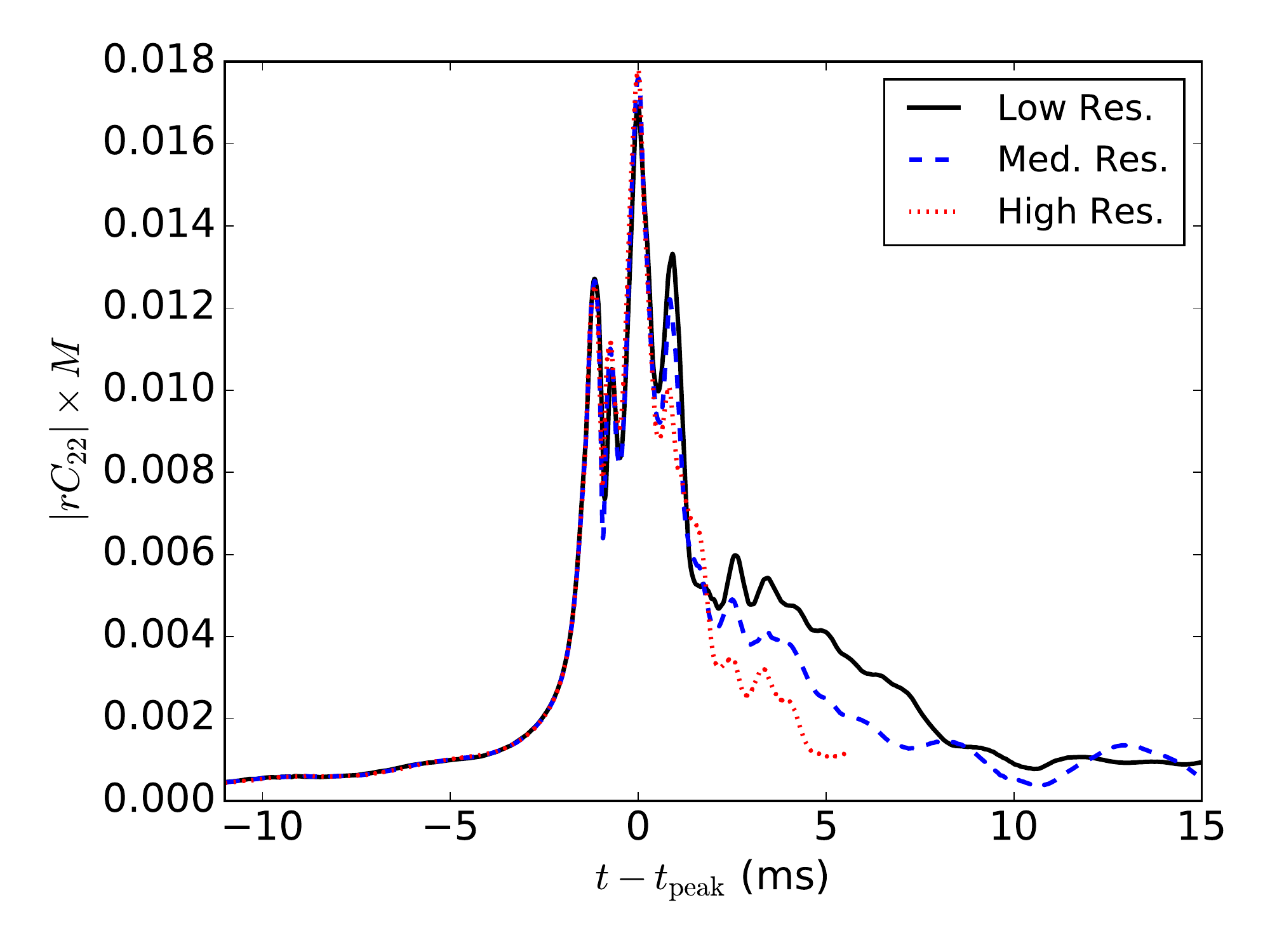}
	\includegraphics[width = 3.5in]{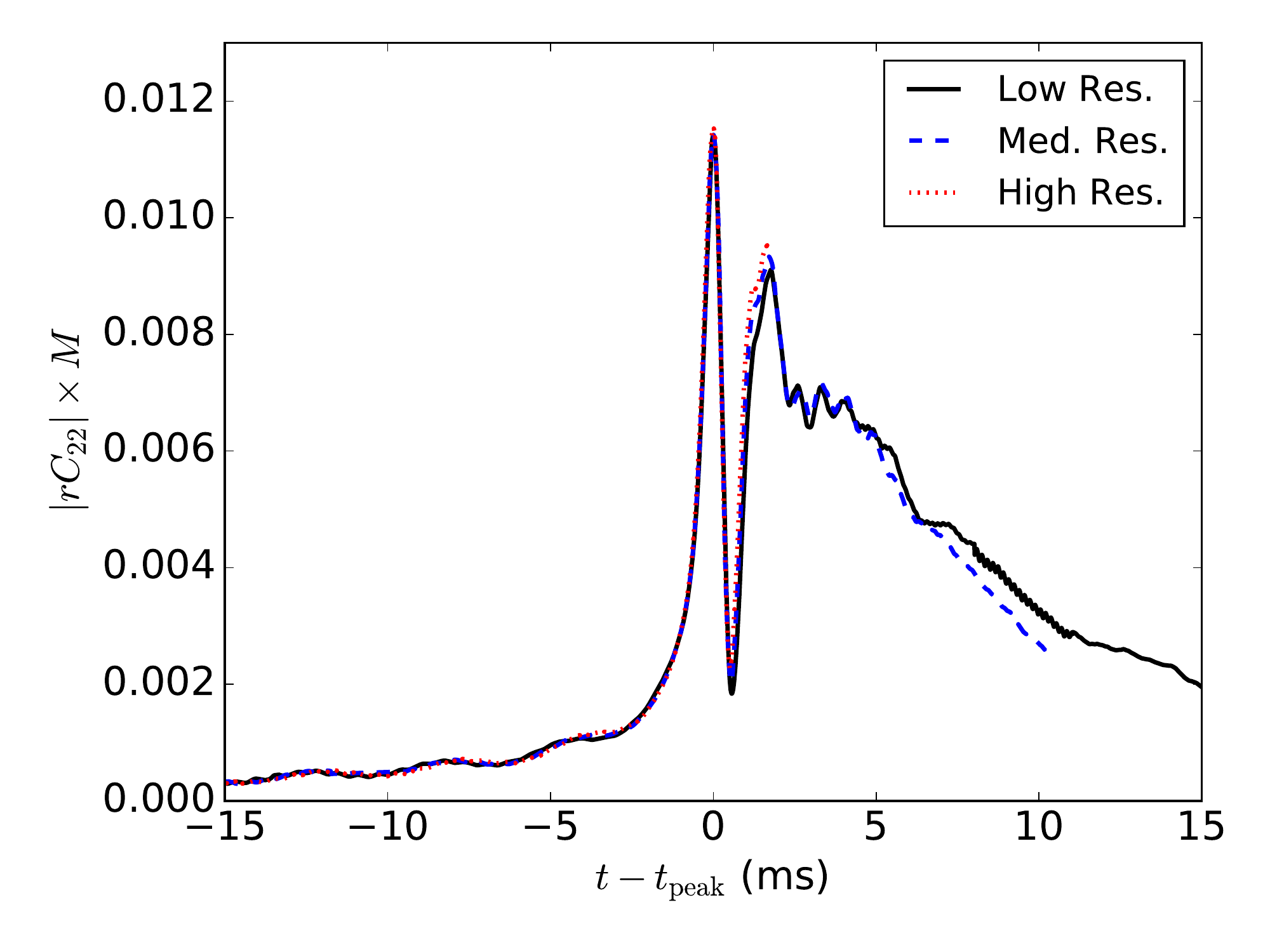}
\caption{ The GW amplitude (magnitude of $\ell=m=2$
  component of $\psi_4$ multiplied by the extraction radius $r$) from mergers with the ENG EOS, with
  nonspinning NSs shown on the left, and the largest value of NS spin
  considered ($a_{\rm NS}=0.33$) shown on the right. Three different
  resolutions are plotted, and the curves have been aligned in time at
  peak amplitude.  The phase error is shown in
  Fig.~\ref{fig:gw_phase_conv}.  } \label{fig:gw_amp_conv}
\end{center}
\end{figure*}

\begin{figure*}
\begin{center}
	\includegraphics[width = 3.5in]{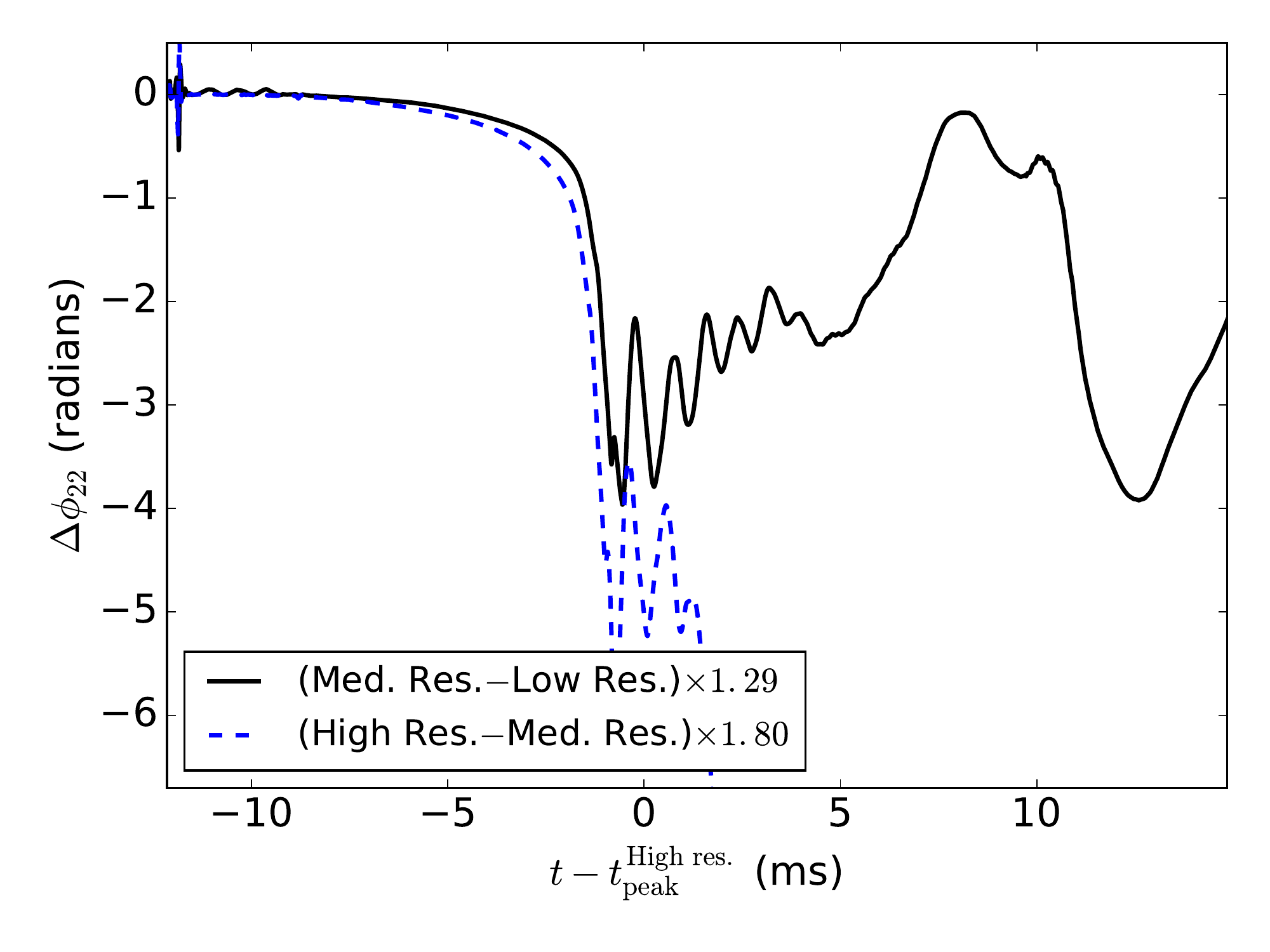}
	\includegraphics[width = 3.5in]{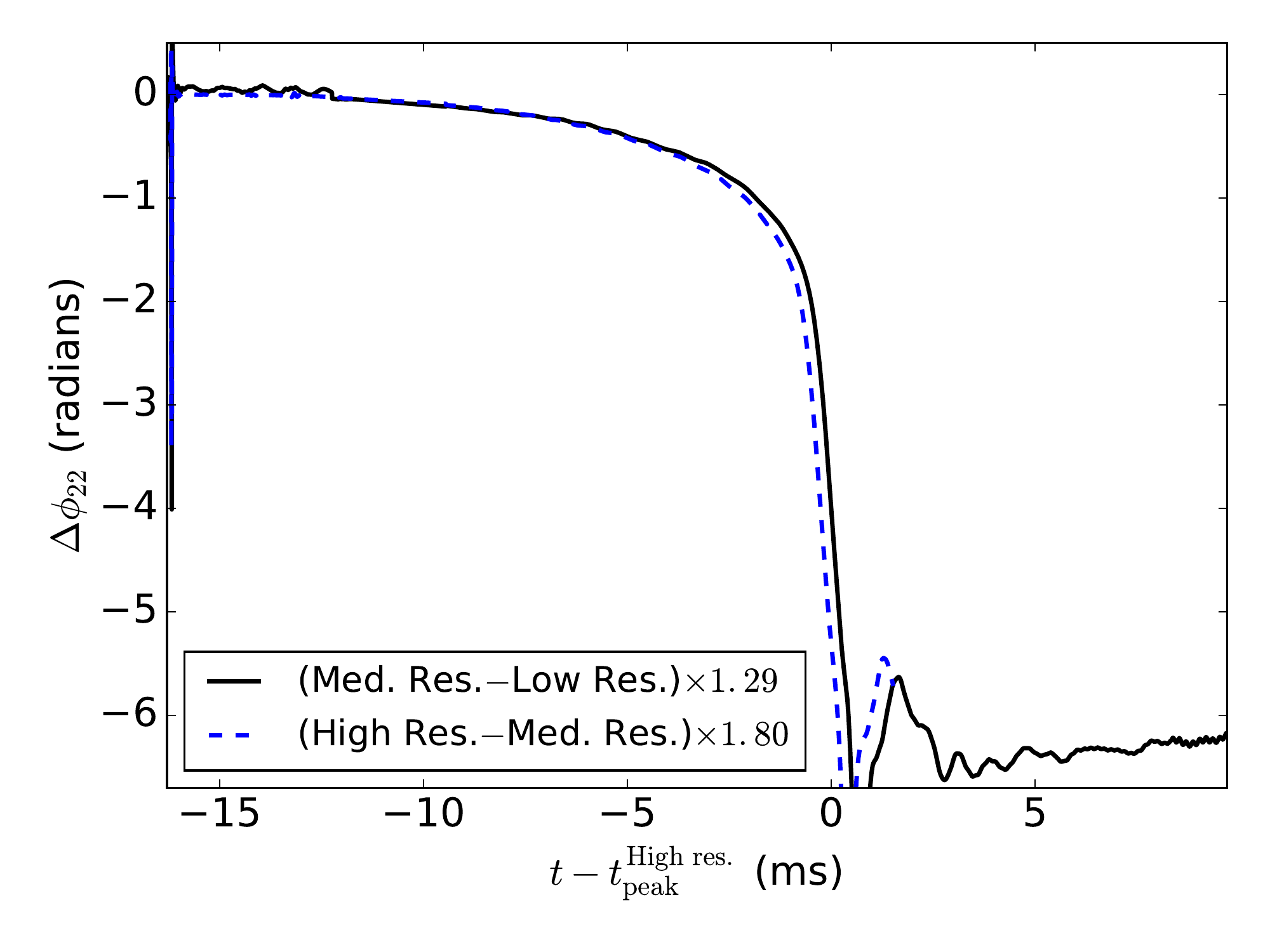}
\caption{ The difference in the GW phase (phase of
  $\ell=m=2$ component of $\psi_4$) across different resolutions from
  mergers with the ENG EOS and nonspinning NSs (left), and the
  largest value of NS spin considered ($a_{\rm NS}=0.33$; right).  The
  differences have been scaled assuming second order
  convergence. 
  Time is shown on the vertical axis with respect to where the peak of the GW
  signal occurs in the highest resolution case (in particular, the GWs have not been aligned at merger).
    } \label{fig:gw_phase_conv}
\end{center}
\end{figure*}

As noted in the main text, for select cases we perform a resolution study
utilizing $4/3\times$ and $2\times$ the default resolution, which has
$dx\approx 0.05 M$ on the finest level of mesh refinement.  In
Figs.~\ref{fig:gw_amp_conv} and~\ref{fig:gw_phase_conv}, we show, respectively,
how the amplitude and phase of the GWs vary with resolution for the merger of
nonspinning NSs and the merger with the highest value of NS spin ($a_{\rm
NS}=0.33$) for the ENG EOS.  In both cases, we can see that leading up to
merger the differences across resolutions are small, though post-merger the
phase errors become significant, as is typically found in BNS simulations.
Examining the resolution dependence of the peak frequency of the post-merger
GWs for these two cases, we find the differences between the two lower
resolutions (the ones we continue for at least 10 ms post-merger) to be small
(on the order of Hz) compared to the differences due to NS spin.

Notice that Fig.~\ref{fig:gw_amp_conv} explicitly demonstrates that the
post-merger GW amplitude for the highest spinning case has a small dependence
on numerical resolution for the parameters used here. By contrast, the
irrotational case has a more significant resolution dependence post-merger.
However, the nonspinning cases show that the higher the resolution, the faster
the post-merger GW amplitude decays. Therefore, we can conclude that the
post-merger GW amplitude in nonzero spin cases decays more slowly in the first
20 ms than in the irrotational case, as stated in the main text.

\begin{figure}
\begin{center}
	\includegraphics[width = 3.5in]{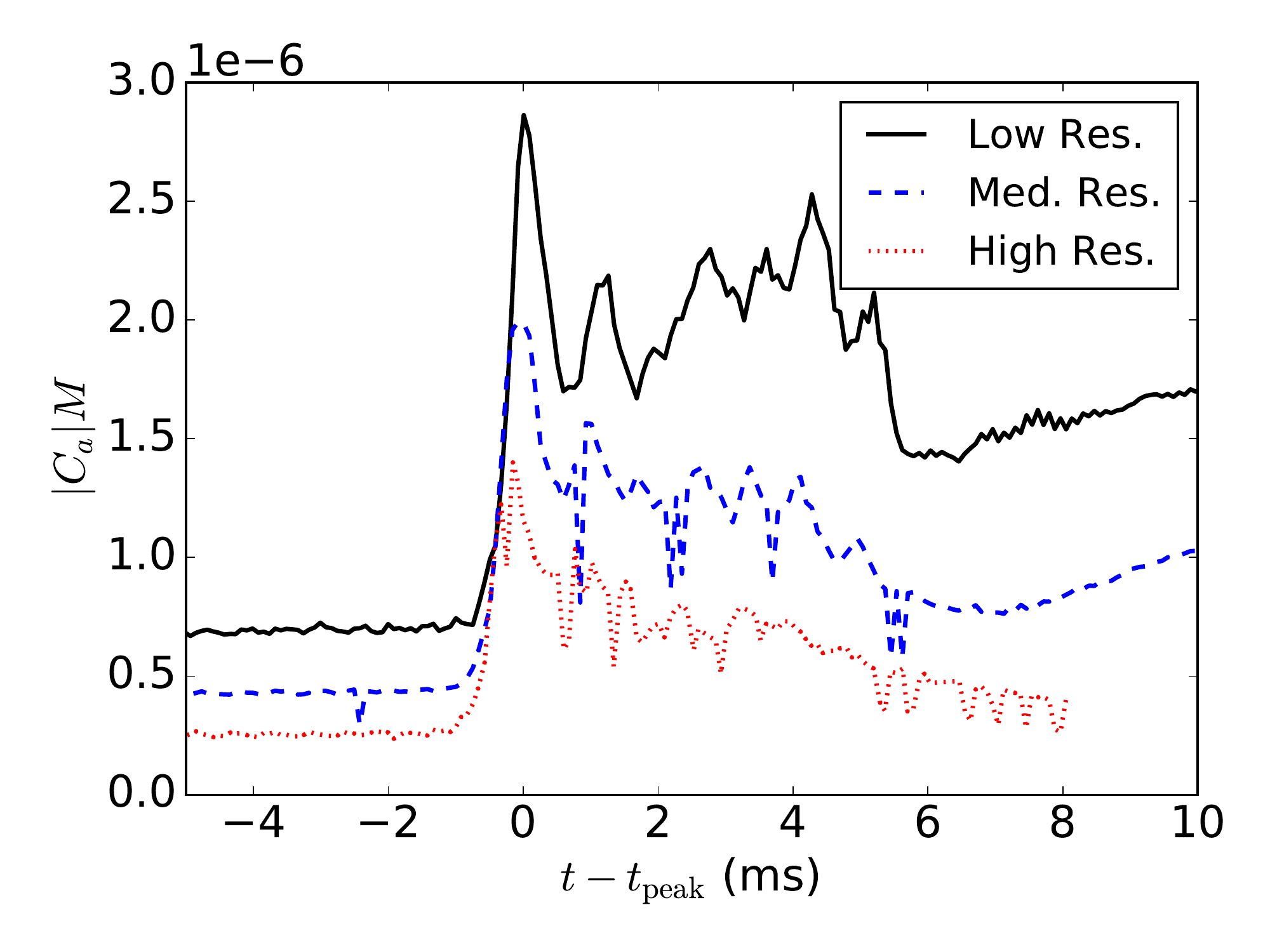}
\caption{
    The norm of the generalized-harmonic constraint $C^a=\Box x^a - H^a$
    integrated in the equatorial plane for several different resolutions
    for the IR, ENG EOS case.
 } \label{fig:cnst_conv}
\end{center}
\end{figure}

We also show the convergence of the constraint violation, leading up
to and post-merger in Fig.~\ref{fig:cnst_conv}. The convergence
towards zero with resolution of this quantity is approximately
consistent with second order, as expected.

\begin{figure*}
\begin{center}
	\includegraphics[width = 3.5in]{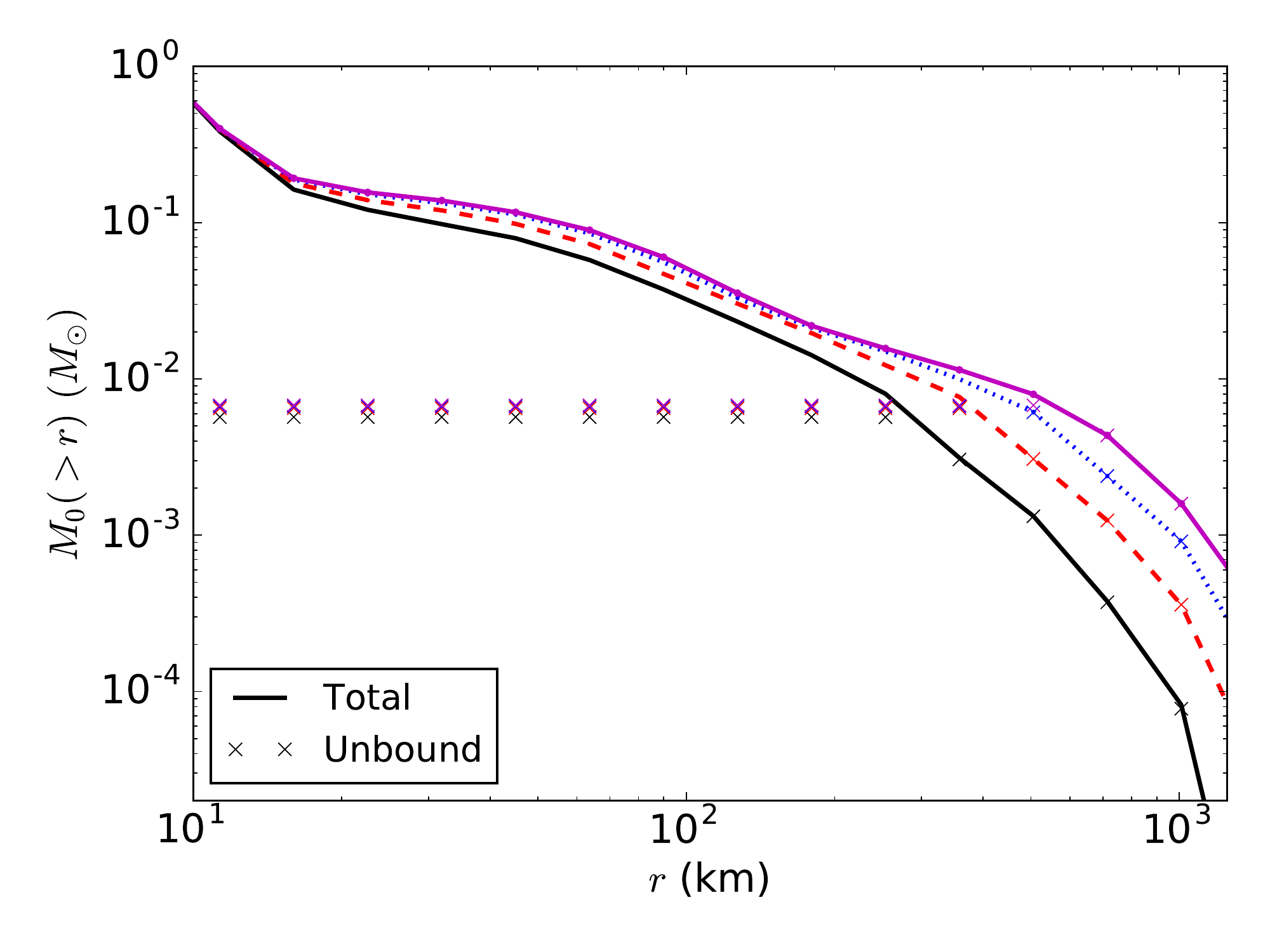}
        \includegraphics[width = 3.5in]{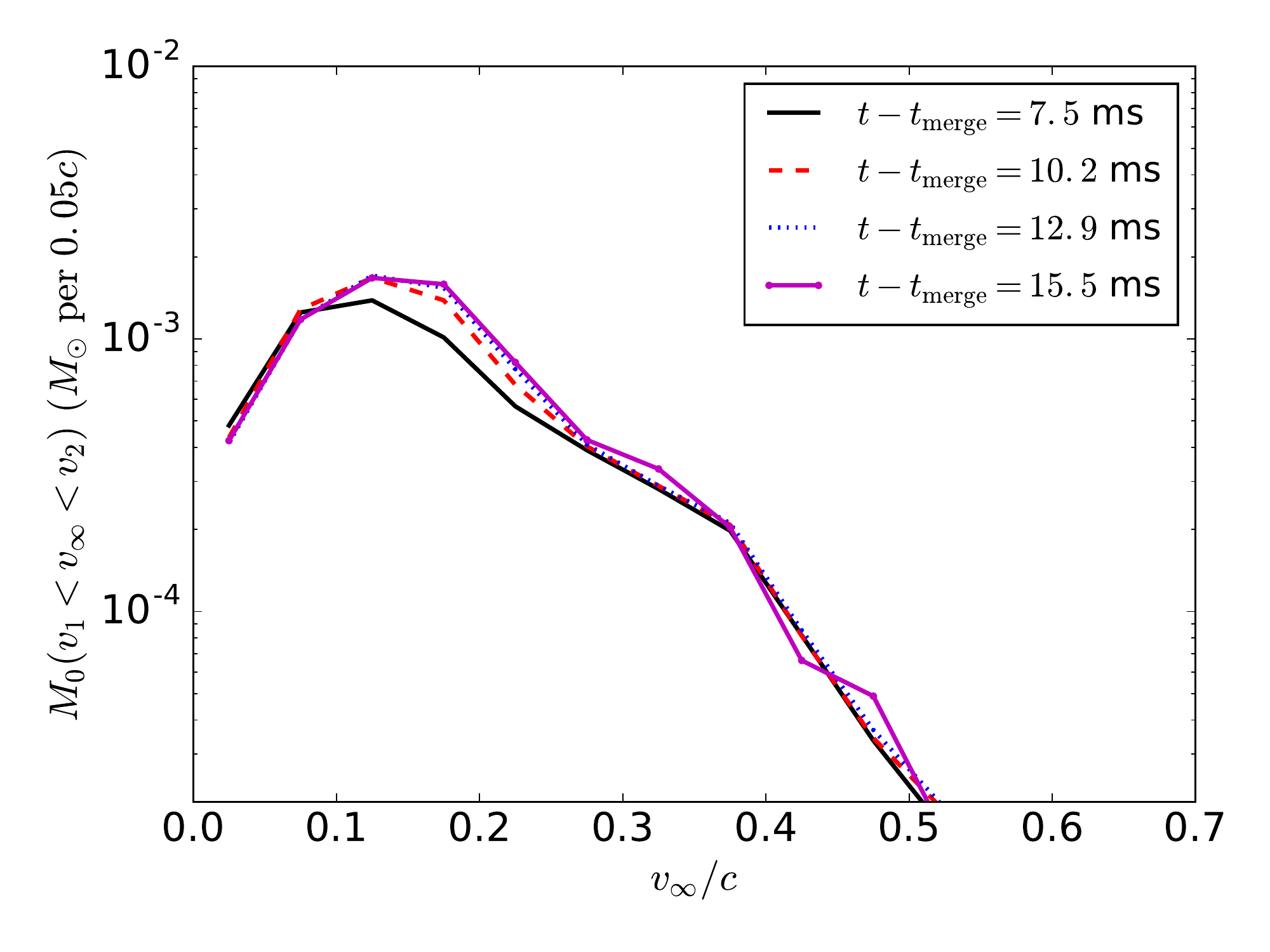}
\caption{ Same as Fig.~\ref{fig:ub_eng}, but just showing the case with the ENG
  EOS and no NS spin at different times post-merger.  The legend from the right panel
  applies to the left panel as well.  } \label{fig:ub_c0}
\end{center}
\end{figure*}

Finally, we comment on the measurement of the bound/unbound matter. 
In Fig.~\ref{fig:ub_c0} we show how this depends on time for an example case.
From this it can be seen that these quantities are relatively constant for
$t-t_{\rm merge}>10$ ms (e.g., the total unbound matter increases by $<5\%$), with the material marked as unbound moving outward in 
radius as expected, while maintaining the same distribution in velocity (at infinity).
The results shown in the main text are from the latest time shown in Fig.~\ref{fig:ub_c0}.

We have not run the highest resolution case sufficiently long to compute
this diagnostic, but comparing the lower two resolutions, we find a relative difference
of $\approx 9\%$ in the unbound material for the ENG EOS nonspinning case, 
and a somewhat larger difference of $30\%$ in the $a_{\rm NS}=0.33$ case.
For reference we note that we use a so-called ``artificial atmosphere,"
as is typical in such hydrodynamical simulations, that has a maximum density
of $\approx 8\times 10^4$ gm/cm$^3$ 
(i.e. roughly 10 orders of magnitude below the maximum density of the NSs)
and gradually decreases in density towards the outer boundary, as described in
Ref.~\cite{code_paper}.

\bibliographystyle{apsrev4-1.bst}
\bibliography{ref}

\end{document}